\pdfoutput=1
\documentclass[12pt,a4paper]{article}

\usepackage{ifthen} 

\newboolean{pdflatex}
\setboolean{pdflatex}{true} 

\newboolean{articletitles}
\setboolean{articletitles}{true} 

\newboolean{uprightparticles}
\setboolean{uprightparticles}{false} 

\def\paperauthors{LHCb collaboration} 
\def\paperasciititle{Measurement of Xi_{c}^{+} production in pPb collisions at sqrt(snn)=8.16 TeV at LHCb} 
\def\papertitle{Measurement of $\Xi_{c}^{+}$ production in $p$Pb collisions at $\sqsnn=8.16$\tev at LHCb} 
\def\paperkeywords{{High Energy Physics}, {LHCb}} 
\def\papercopyright{\the\year\ CERN for the benefit of the LHCb collaboration} 
\def\paperlicence{CC BY 4.0 licence}
\def\paperlicenceurl{https://creativecommons.org/licenses/by/4.0/}

\newcommand{\eacc}{{\ensuremath{\varepsilon_{\mathrm{ acc}}}}\xspace}
\newcommand{\epid}{{\ensuremath{\varepsilon_{\mathrm{ PID}}}}\xspace}

\usepackage[top=1in, bottom=1.25in, left=1in, right=1in]{geometry}

\usepackage{microtype}
\usepackage{lineno}  
\usepackage{xspace} 
\usepackage{caption} 

\usepackage{graphicx}  
\usepackage{color}
\usepackage{colortbl}
\graphicspath{{./figs/}} 

\usepackage{amsmath} 
\usepackage{amssymb}
\usepackage{amsfonts}
\usepackage{upgreek} 

\newcommand*\patchAmsMathEnvironmentForLineno[1]{%
\expandafter\let\csname old#1\expandafter\endcsname\csname #1\endcsname
\expandafter\let\csname oldend#1\expandafter\endcsname\csname
end#1\endcsname
 \renewenvironment{#1}%
   {\linenomath\csname old#1\endcsname}%
   {\csname oldend#1\endcsname\endlinenomath}%
}
\newcommand*\patchBothAmsMathEnvironmentsForLineno[1]{%
  \patchAmsMathEnvironmentForLineno{#1}%
  \patchAmsMathEnvironmentForLineno{#1*}%
}
\AtBeginDocument{%
\patchBothAmsMathEnvironmentsForLineno{equation}%
\patchBothAmsMathEnvironmentsForLineno{align}%
\patchBothAmsMathEnvironmentsForLineno{flalign}%
\patchBothAmsMathEnvironmentsForLineno{alignat}%
\patchBothAmsMathEnvironmentsForLineno{gather}%
\patchBothAmsMathEnvironmentsForLineno{multline}%
\patchBothAmsMathEnvironmentsForLineno{eqnarray}%
}

\usepackage{hyperxmp}

\usepackage[pdftex,
            pdfauthor={\paperauthors},
            pdftitle={\paperasciititle},
            pdfkeywords={\paperkeywords},
            pdfcopyright={Copyright (C) \papercopyright},
            pdflicenseurl={\paperlicenceurl}]{hyperref}

\usepackage[colorinlistoftodos,textsize=scriptsize]{todonotes}

\usepackage[bottom,flushmargin,hang,multiple]{footmisc}

\usepackage[all]{hypcap} 

\usepackage{xspace} 
\usepackage{upgreek}

\def\lhcb   {\mbox{LHCb}\xspace}

\def\MagUp {\mbox{\em Mag\kern -0.05em Up}\xspace}

\ifthenelse{\boolean{uprightparticles}}%
{

 \def\Ppi         {\ensuremath{\uppi}\xspace}

 \def\PDelta      {\ensuremath{\Delta}\xspace}                 
 \def\PXi         {\ensuremath{\Xi}\xspace}                 
 \def\PLambda     {\ensuremath{\Lambda}\xspace}                 
 \def\PSigma      {\ensuremath{\Sigma}\xspace}                 
 \def\POmega      {\ensuremath{\Omega}\xspace}                 
 \def\PUpsilon    {\ensuremath{\Upsilon}\xspace}
 \let\oldPi\Pi
 \def\PPi         {\ensuremath{\oldPi}\xspace}

 \def\PB      {\ensuremath{\mathrm{B}}\xspace}                 
                  
 \def\PD      {\ensuremath{\mathrm{D}}\xspace}

 \def\PK      {\ensuremath{\mathrm{K}}\xspace}

 \def\Pb      {\ensuremath{\mathrm{b}}\xspace}                 
 \def\Pc      {\ensuremath{\mathrm{c}}\xspace}                 
                  
 \def\Pe      {\ensuremath{\mathrm{e}}\xspace}

 \def\Pi      {\ensuremath{\mathrm{i}}\xspace}

 \def\Pp      {\ensuremath{\mathrm{p}}\xspace}

 \def\Ps      {\ensuremath{\mathrm{s}}\xspace}

 \def\thebaroffset{0.0em}
}
{

 \def\Ppi         {\ensuremath{\pi}\xspace}

 \mathchardef\PDelta="7101
 \mathchardef\PXi="7104
 \mathchardef\PLambda="7103
 \mathchardef\PSigma="7106
 \mathchardef\POmega="710A
 \mathchardef\PUpsilon="7107
 \mathchardef\PPi="7105
                  
 \def\PB      {\ensuremath{B}\xspace}                 
                  
 \def\PD      {\ensuremath{D}\xspace}

 \def\PK      {\ensuremath{K}\xspace}

 \def\Pb      {\ensuremath{b}\xspace}                 
 \def\Pc      {\ensuremath{c}\xspace}                 
                  
 \def\Pe      {\ensuremath{e}\xspace}

 \def\Pi      {\ensuremath{i}\xspace}

 \def\Pp      {\ensuremath{p}\xspace}

 \def\Ps      {\ensuremath{s}\xspace}

 \def\thebaroffset{0.18em}
}
\newcommand{\offsetoverline}[2][\thebaroffset]{\kern #1\overline{\kern -#1 #2}}%

\makeatletter
\ifcase \@ptsize \relax
  \newcommand{\miniscule}{\@setfontsize\miniscule{4}{5}}
\or
  \newcommand{\miniscule}{\@setfontsize\miniscule{5}{6}}
\or
  \newcommand{\miniscule}{\@setfontsize\miniscule{5}{6}}
\fi
\makeatother

\DeclareRobustCommand{\optbar}[1]{\shortstack{{\miniscule (\rule[.5ex]{1.25em}{.18mm})}
  \\ [-.7ex] $#1$}}

\def\electron   {{\ensuremath{\Pe}}\xspace}
\def\ep         {{\ensuremath{\Pe^+}}\xspace}

\def\squark    {{\ensuremath{\Ps}}\xspace}

\def\cquark    {{\ensuremath{\Pc}}\xspace}

\def\bquark    {{\ensuremath{\Pb}}\xspace}

\def\pion   {{\ensuremath{\Ppi}}\xspace}

\def\pip    {{\ensuremath{\pion^+}}\xspace}

\def\kaon    {{\ensuremath{\PK}}\xspace}

\def\KorKbar {\kern \thebaroffset\optbar{\kern -\thebaroffset \PK}{}\xspace}

\def\Km      {{\ensuremath{\kaon^-}}\xspace}

\def\D       {{\ensuremath{\PD}}\xspace}

\def\DorDbar {\kern \thebaroffset\optbar{\kern -\thebaroffset \PD}\xspace}

\def\Dp      {{\ensuremath{\D^+}}\xspace}
\def\Dm      {{\ensuremath{\D^-}}\xspace}

\def\DpDm    {\ensuremath{\Dp {\kern -0.16em \Dm}}\xspace}

\def\B       {{\ensuremath{\PB}}\xspace}

\def\BorBbar {\kern \thebaroffset\optbar{\kern -\thebaroffset \PB}\xspace}

\def\Bd      {{\ensuremath{\B^0}}\xspace}

\def\BdorBdbar {\kern \thebaroffset\optbar{\kern -\thebaroffset \Bd}\xspace}

\def\Bs      {{\ensuremath{\B^0_\squark}}\xspace}

\def\BsorBsbar {\kern \thebaroffset\optbar{\kern -\thebaroffset \Bs}\xspace}

\def\Y#1S{\ensuremath{\PUpsilon{(#1S)}}\xspace}

\def\proton      {{\ensuremath{\Pp}}\xspace}

\def\Lz          {{\ensuremath{\PLambda}}\xspace}

\def\LorLbar     {\kern \thebaroffset\optbar{\kern -\thebaroffset \PLambda}\xspace}

\def\Xires       {{\ensuremath{\PXi}}\xspace}

\def\Lc          {{\ensuremath{\Lz^+_\cquark}}\xspace}

\def\Xicz        {{\ensuremath{\Xires^0_\cquark}}\xspace}
\def\Xicp        {{\ensuremath{\Xires^+_\cquark}}\xspace}

\def\Lb           {{\ensuremath{\Lz^0_\bquark}}\xspace}

\newcommand{\esel}{{\ensuremath{\varepsilon_{\mathrm{ sel/rec}}}}\xspace}
\newcommand{\etrg}{{\ensuremath{\varepsilon_{\mathrm{ trg/sel}}}}\xspace}

\def\AT#1     {\ensuremath{A_{\mathrm{T}}^{#1}}\xspace}           

\def\C#1      {\ensuremath{\mathcal{C}_{#1}}\xspace}                       
\def\Cp#1     {\ensuremath{\mathcal{C}_{#1}^{'}}\xspace}                    
\def\Ceff#1   {\ensuremath{\mathcal{C}_{#1}^{\mathrm{(eff)}}}\xspace}        
\def\Cpeff#1  {\ensuremath{\mathcal{C}_{#1}^{'\mathrm{(eff)}}}\xspace}       
\def\Ope#1    {\ensuremath{\mathcal{O}_{#1}}\xspace}                       
\def\Opep#1   {\ensuremath{\mathcal{O}_{#1}^{'}}\xspace}                    

\newcommand{\nospaceunit}[1]{\ensuremath{\text{#1}}}       
\newcommand{\aunit}[1]{\ensuremath{\text{\,#1}}}       

\newcommand{\tev}{\aunit{Te\kern -0.1em V}\xspace}
\newcommand{\gev}{\aunit{Ge\kern -0.1em V}\xspace}
\newcommand{\mev}{\aunit{Me\kern -0.1em V}\xspace}
\newcommand{\kev}{\aunit{ke\kern -0.1em V}\xspace}
\newcommand{\ev}{\aunit{e\kern -0.1em V}\xspace}
 
\newcommand{\mevc}{\ensuremath{\aunit{Me\kern -0.1em V\!/}c}\xspace}
\newcommand{\gevc}{\ensuremath{\aunit{Ge\kern -0.1em V\!/}c}\xspace}
\newcommand{\mevcc}{\ensuremath{\aunit{Me\kern -0.1em V\!/}c^2}\xspace}
\newcommand{\gevcc}{\ensuremath{\aunit{Ge\kern -0.1em V\!/}c^2}\xspace}

\def\mum  {\ensuremath{\,\upmu\nospaceunit{m}}\xspace}

\def\nb {\aunit{nb}\xspace}
\def\invnb {\ensuremath{\nb^{-1}}\xspace}

\newcommand{\chisqip}{\ensuremath{\chi^2_{\text{IP}}}\xspace}

\def\gsim{{~\raise.15em\hbox{$>$}\kern-.85em
          \lower.35em\hbox{$\sim$}~}\xspace}
\def\lsim{{~\raise.15em\hbox{$<$}\kern-.85em
          \lower.35em\hbox{$\sim$}~}\xspace}

\def\sPlot{\mbox{\em sPlot}\xspace}

\def\sqs   {\ensuremath{\protect\sqrt{s}}\xspace}
\def\sqsnn {\ensuremath{\protect\sqrt{s_{\scriptscriptstyle{NN}}}}\xspace}
\def\pt         {\ensuremath{p_{\mathrm{T}}}\xspace}

\def\ptot       {\ensuremath{p}\xspace}

\def\evtgen     {\mbox{\textsc{EvtGen}}\xspace}

\def\geant      {\mbox{\textsc{Geant4}}\xspace}

\def\pythia     {\mbox{\textsc{Pythia}}\xspace}

\def\tell1  {TELL1\xspace}
\def\ukl1   {UKL1\xspace}

\newcommand{\lhcborcid}[1]{\href{https://orcid.org/#1}{\hspace*{0.1em}\raisebox{-0.45ex}{\includegraphics[width=1em]{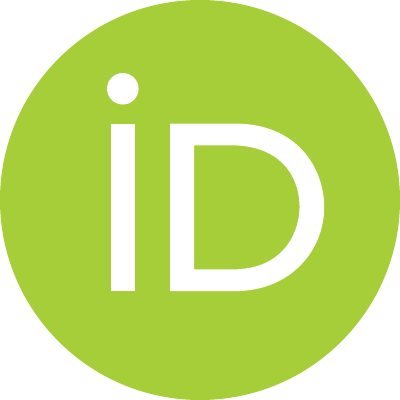}}}}

\usepackage{cite} 
\usepackage{mciteplus}

\usepackage{longtable} 
\begin{document}

\renewcommand{\thefootnote}{\fnsymbol{footnote}}
\setcounter{footnote}{1}


\begin{titlepage}
\pagenumbering{roman}

\vspace*{-1.5cm}
\centerline{\large EUROPEAN ORGANIZATION FOR NUCLEAR RESEARCH (CERN)}
\vspace*{1.5cm}
\noindent
\begin{tabular*}{\linewidth}{lc@{\extracolsep{\fill}}r@{\extracolsep{0pt}}}
\ifthenelse{\boolean{pdflatex}}
{\vspace*{-1.5cm}\mbox{\!\!\!\includegraphics[width=.14\textwidth]{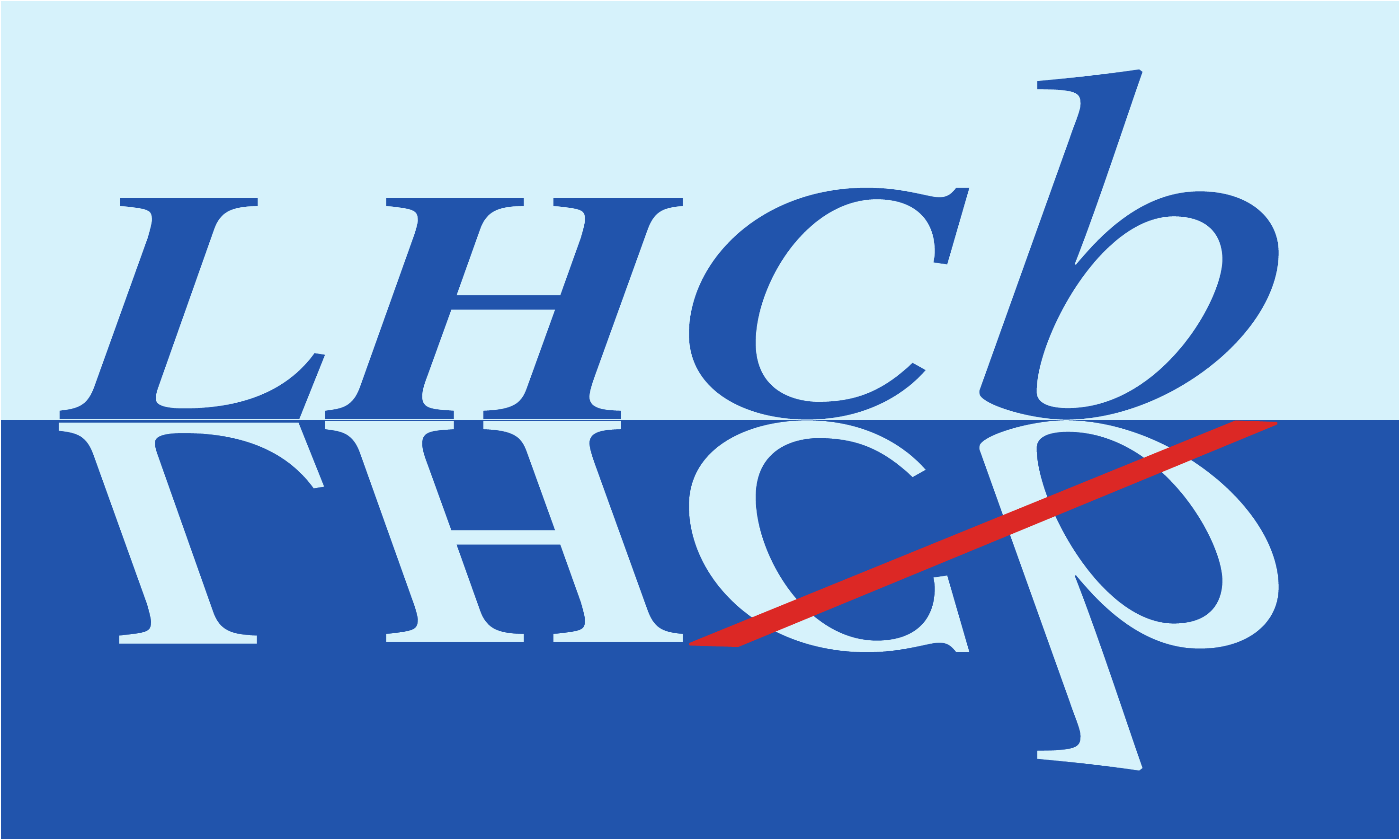}} & &}%
{\vspace*{-1.2cm}\mbox{\!\!\!\includegraphics[width=.12\textwidth]{figs/lhcb-logo.eps}} & &}%
\\
 & & CERN-EP-2023-057 \\  
 & & LHCb-PAPER-2022-041 \\  
 & & 2 November 2023 \\ 
 & & \\
\end{tabular*}

\vspace*{4.0cm}

{\normalfont\bfseries\boldmath\huge
\begin{center}
  \papertitle 
\end{center}
}

\vspace*{2.0cm}

\begin{center}
\paperauthors\footnote{Authors are listed at the end of this paper.}
\end{center}

\vspace{\fill}

\begin{abstract}
  \noindent
  A study of prompt \Xicp production in proton-lead collisions is performed with the LHCb experiment at a center-of-mass energy per nucleon pair of 8.16~\tev in 2016 in $p$Pb and Pb$p$ collisions with an estimated integrated luminosity of approximately 12.5 and 17.4~\invnb, respectively. The \Xicp production cross section, as well as the \Xicp to \Lc production cross-section ratio, are measured as a function of the transverse momentum and rapidity and compared to the latest theory predictions. The forward-backward asymmetry is also measured as a function of the \Xicp\ transverse momentum. The results provide strong constraints on theoretical calculation and are a unique input for hadronization studies in different collision systems.
  
\end{abstract}

\vspace*{2.0cm}

\begin{center}
  Published in
  Phys.~Rev.~C 109, 044901
\end{center}

\vspace{\fill}

{\footnotesize 
\centerline{\copyright~\papercopyright. \href{\paperlicenceurl}{\paperlicence}.}}
\vspace*{2mm}

\end{titlepage}


\newpage
\setcounter{page}{2}
\mbox{~}
%
%
%
%


\renewcommand{\thefootnote}{\arabic{footnote}}
\setcounter{footnote}{0}

\cleardoublepage


\pagestyle{plain} 
\setcounter{page}{1}
\pagenumbering{arabic}


\section{Introduction}

In hadronic collisions, heavy quarks are produced in hard scattering processes with large momentum transfer. Theoretical predictions based on perturbative quantum chromo-dynamics (pQCD) describe reasonably well the transverse momentum (\pt) differential charm production cross sections in proton-proton ($pp$) collisions at different energies~\cite{LHCb-PAPER-2016-042}. The heavy-flavor hadron cross sections are usually computed using the factorization approach as a convolution of three terms~\cite{Collins:1985gm}: the parton distribution functions of the colliding particles, the hard-scattering cross section, and the fragmentation function (FF) of heavy quarks into a given heavy-flavor hadron. It is assumed that the FFs are universal between collision systems and energies. At the CERN Large Hadron Collider (LHC), the FFs used to describe the measurements of heavy-flavor hadron production in $pp$ collisions at different center-of-mass energies are tuned on \ep\electron$^-$ collision data within the framework of pQCD over a wide \pt range. However, evidence of nonuniversal FFs was observed by the LHCb collaboration in a study of the $\Lb$ baryon to $B^{-}$ and $\bar{B}^{0}$ meson production ratio in $pp$ collisions~\cite{LHCb-PAPER-2011-018}. Multiple measurements of the relative production of different heavy-flavor hadron species have been made, as they are sensitive probes of FFs.

The measurements of heavy-flavored hadron ratios in hadronic collisions, such as the \Lc/$D^{0}$ cross-section ratio, in $pp$ collisions at \sqs = 5, 7 and 13 \tev and in proton-lead ($p$Pb) collisions at \sqs = 5 \tev~\cite{Werner:2023fne,ALICE:2017thy,LHCb-PAPER-2018-021}, show an enhancement with respect to predictions from pQCD calculations with charm fragmentation based on $e^+e^-$~\cite{ARGUS:1991vjh, CLEO:1990unu} and $e^-p$~\cite{ZEUS:2005pvv, ZEUS:2010cic, Braaten:1994bz} measurements. Similar observations were made in the measurement of the \Xicz (\Xicp)/$D^{0}$ cross section ratio in $pp$ collisions at \sqs = 7 and 13 \tev~\cite{ALICE:2017dja, ALICE:2021bli}.
Multiple models explain the ratio enhancement. For example, color reconnection~\cite{Christiansen:2015yqa} could be stronger in $pp$ collisions than in $e^+e^-$ collisions, resulting in an enhanced production of baryons relative to mesons. Other models predicts that hadronization via coalescence~\cite{Greco:2004rm} will take place. 

This paper presents a study of the ratio of production of the \Xicp particle, a charm-strange baryon, to the production of the \Lc baryon in $p$Pb collisions at the LHCb experiment. This measurement has the potential to shed light on the mechanism of hadronization and its universality as it is the first one to be performed in proton-nucleus collisions, considered the best environment to study 
 the so-called cold nuclear matter (CNM) effects, such as shadowing~\cite{Armesto:2006ph, Vitev:2007ve, PhysRevLett.68.1834}, energy loss~\cite{dai2021parton} and nuclear break-up~\cite{breakup2008}. In nucleus-nucleus collisions, in addition to CNM effects, experimental results indicate the formation of a high-density color-deconfined medium, the quark-gluon plasma (QGP), a state of matter with asymptotically free partons, which is expected to exist at extremely high temperature and density. The presence of QGP can be determined by observing the change in behavior of the particles as they traverse the nuclear medium with respect to their behavior in the absence of QGP~\cite{2013aliceheavy}. In $p$Pb collisions the energy density is not expected to be sufficient to produce a QGP medium; however, some theoretical models predict the formation of ``QGP droplets'' \cite{Lee:2018njd}, which could partially induce in $p$Pb the same behavior, albeit less pronounced, as in PbPb collisions. Moreover, several QGP models predict that strange ($s$) quark production is enhanced in heavy-ion collisions, as first reported in Ref.~\cite{RAFELSKI1984215}; thus, strangeness enhancement and strange antibaryon abundance are considered signatures for QGP formation~\cite{Rafelski:1980fy,Koch:1984tz,Koch:1986ud}, mainly due to the predominance of the gluonic production mechanism $gg\rightarrow s\bar{s}$. Strangeness enhancement is investigated at accelerators by studying the ratio of production rates of hadrons containing a strange quark to those without, as done, by experiments at CERN~\cite{2017nature} and BNL Relativistic Heavy Ion Collider (RHIC)~\cite{PHENIX:2013kod,STAR:2006uve}. The ratio of  \Xicp to \Lc production measured in this paper, by comparing a baryon with strange content to one without, has the potential to test the presence of QGP formation from QGP droplets in $p$Pb collisions, which has not yet been established.
\section{Analysis, detector, and simulation}

 The \Xicp and \Lc candidates\footnote{Charge conjugate decays are implied throughout this paper, unless otherwise stated.} in the analysis are reconstructed via the hadronic decay to the $\proton\Km\pip$ final state. The measurements are performed as a function of \pt and rapidity ($y^*$) of the baryons, using $p$Pb and Pb$p$ collisions with an estimated integrated luminosity of approximately 12.5 and 17.4~\invnb~\cite{LHCb-PAPER-2014-047}, respectively, collected by the LHCb detector at center-of-mass energy per nucleon pair of $\sqsnn=8.16\tev$. The rapidity $y^*$ in the nucleon-nucleon center-of-mass system is related to the rapidity in the laboratory frame ($y_{\textrm{lab}}$) via $y^* = y_{\textrm{lab}} \pm 0.4645$. 
 Here, 0.4645 is the shift in rapidity in the direction of the proton beam due to the unequal mass of the two colliding objects. The proton beam and the lead beam have different energies per nucleon in the laboratory frame, with $E_{p}$~=~6.5\tev and $E_{\textrm{Pb}}$~=~2.56\tev. The particles are separated according to whether they originate from the collision point (prompt) or from the decay of $b$ hadrons (nonprompt) using the impact parameter (IP), defined as the distance of closest approach of the particle trajectory to the collision point. Data are analyzed separately in the $p$Pb (forward) sample, covering a rapidity range 1.5~$<y^*<$~4.0, and the Pb$p$ (backward) sample, covering the range $-$5.0~$<y^*<$~$-$2.5.

The double differential cross section for prompt \Xicp (\Lc) production is measured as
\begin{align}	
	\frac{\textrm{d}^{2}\sigma_{\Xicp(\Lc)}(\pt, y^{*})}{\textrm{d}\pt\textrm{d}y^{*}} = \frac{N^{\Xicp (\Lc)}(\pt, y^{*})}{\mathcal{L}\cdot \mathcal{B} \cdot \epsilon_{\textrm{tot}}(\pt, y^{*}) \cdot \Delta\pt\Delta y^{*}},
\label{eq1}
\end{align}
where $N^{\Xicp (\Lc)}(\pt, y^{*})$ is the measured signal yield of prompt \Xicp (\Lc) decays produced in a given interval of \pt and $y^{*}$, $\Delta\pt$ and $\Delta y^{*}$, respectively, and $\mathcal{L}$ represents the integrated luminosity. The branching fractions $\mathcal{B}$ are $(0.62\pm0.30)\%$ and $(6.28\pm0.32)\%$, for the decays \mbox{$\Xicp \rightarrow \proton\Km\pip$} and \mbox{$\Lc\rightarrow \proton\Km\pip$}, respectively~\cite{PDG2022}. Finally, $\epsilon_{\textrm{tot}}(\pt, y^{*})$ stands for the total signal efficiency determined in the $\Delta\pt$ and $\Delta y^{*}$ interval.
The production ratio of \Xicp to \Lc, $R_{\Xicp/\Lc}(\pt, y^{*})$, is defined as
\begin{align}	
	R_{\Xicp/\Lc}(\pt, y^{*}) \equiv \frac{ \textrm{d}^{2}\sigma_{\Xicp}(\pt, y^{*})/\textrm{d}\pt\textrm{d}y^{*}}
	                { \textrm{d}^{2}\sigma_{\Lc}(\pt, y^{*})/\textrm{d}\pt\textrm{d}y^{*} }.
\end{align}
The forward-backward asymmetry, $R_{\textrm{FB}}(\pt, y^{*})$, is measured in the overlapping rapidity range $2.5 < |y^{*}| < 4.0$ and is defined as
\begin{align}	
	R_{\textrm{FB}}(\pt, y^{*}) \equiv \frac{ \textrm{d}^{2}\sigma_{\Xicp}(\pt, +|y^{*}|)/\textrm{d}\pt\textrm{d}y^{*}}
	                { \textrm{d}^{2}\sigma_{\Xicp}(\pt, -|y^{*}|)/\textrm{d}\pt\textrm{d}y^{*} }.\label{eq:3}
\end{align}


The \lhcb detector~\cite{LHCb-DP-2008-001,LHCb-DP-2014-002} is a single-arm forward
spectrometer covering the \mbox{pseudorapidity} range $2<\eta <5$,
designed for the study of particles containing \bquark or \cquark
quarks. The detector includes a high-precision tracking system
consisting of a silicon-strip vertex detector surrounding the $pp$
interaction region~\cite{LHCb-DP-2014-001}, a large-area silicon-strip detector located
upstream of a dipole magnet with a bending power of about 
$4{\mathrm{\,Tm}}$, and three stations of silicon-strip detectors and straw
drift tubes~\cite{LHCb-DP-2017-001} placed downstream of the magnet.
The tracking system provides a measurement of the momentum, \ptot, of charged particles with
a relative uncertainty that varies from 0.5\% at low momentum to 1.0\% at 200\gevc.
The minimum distance of a track to a primary $pp$ collision vertex (PV), the impact parameter (IP), 
is measured with a resolution of $(15+29/\pt)\mum$,
where \pt is the component of the momentum transverse to the beam, in\,\gevc.
The different types of charged hadrons, such as the kaons and pions used in this analysis, are distinguished using information
from two ring-imaging Cherenkov detectors~\cite{LHCb-DP-2012-003}. 
Photons, electrons and hadrons are identified by a calorimeter system consisting of
scintillating-pad and preshower detectors, an electromagnetic
and a hadronic calorimeter. Muons are identified by a
system composed of alternating layers of iron and multiwire
proportional chambers~\cite{LHCb-DP-2012-002}.
The online event selection is performed by a trigger~\cite{LHCb-DP-2012-004}, 
which consists of a hardware stage, based on information from the calorimeter and muon
systems, followed by a software stage, which applies a full event
reconstruction.


Simulation is used to model the reconstruction efficiency and the effects of the selection requirements.
Charmed baryons are generated in $pp$ collisions at $\sqs = 8.16$\tev using the \pythia~\cite{Sjostrand:2007gs} generator and are embedded into the \mbox{\textsc{EPOS}}\xspace~\cite{EPOS} generator, which simulates the environment of the proton-lead collision. Particle decays are described by \evtgen~\cite{Lange:2001uf}, while the interaction of particles with the detector, and its response in simulation, are implemented using the \geant toolkit~\cite{Agostinelli:2002hh, Allison:2006ve}.


\section{Candidate selection}

The online event selection is performed by a trigger, which consists of a hardware stage followed by a two-level software stage. In between the two software stages, an alignment and calibration of the detector is performed in near real-time and their results are used in the trigger~\cite{LHCb-PROC-2015-011}. The same alignment and calibration information is propagated to the offline reconstruction, ensuring consistent and high-quality particle identification (PID) information between the trigger and offline software. The identical performance of the online and offline reconstruction offers the opportunity to perform physics analyses directly using candidates reconstructed in the trigger~\cite{LHCb-DP-2012-004,LHCb-DP-2016-001} which the present analysis exploits.
The \Xicp and \Lc baryons are reconstructed by combining three tracks identified as proton, kaon and pion candidates. All the charged tracks are required to be well reconstructed and to have transverse momentum $\pt >$ 400 MeV/$c$ and be incompatible with originating at the PV. The tracks must also be within the LHCb acceptance 2.0 $< \eta <$ 5.0 and in the kinematic range 3.2 $< p <$ 100.0 \gevc. The \Xicp and \Lc candidates are required to have a good quality secondary vertex and a reconstructed decay time between 0.1 and 10 ps. 
The angle between the reconstructed candidate momentum and  the vector pointing from the PV to the secondary vertex is required to be close to zero. The $\proton\Km\pip$ invariant mass is required to be within $\pm$70 $\mevcc$ of the known \Xicp(\Lc) mass~\cite{PDG2022}. The invariant mass resolution for the reconstructed candidates is around 10 \mevcc.


\section{Determination of signal yields and efficiency}

\begin{figure*}[b!]
  \begin{center}
    \includegraphics[width=0.46\linewidth]{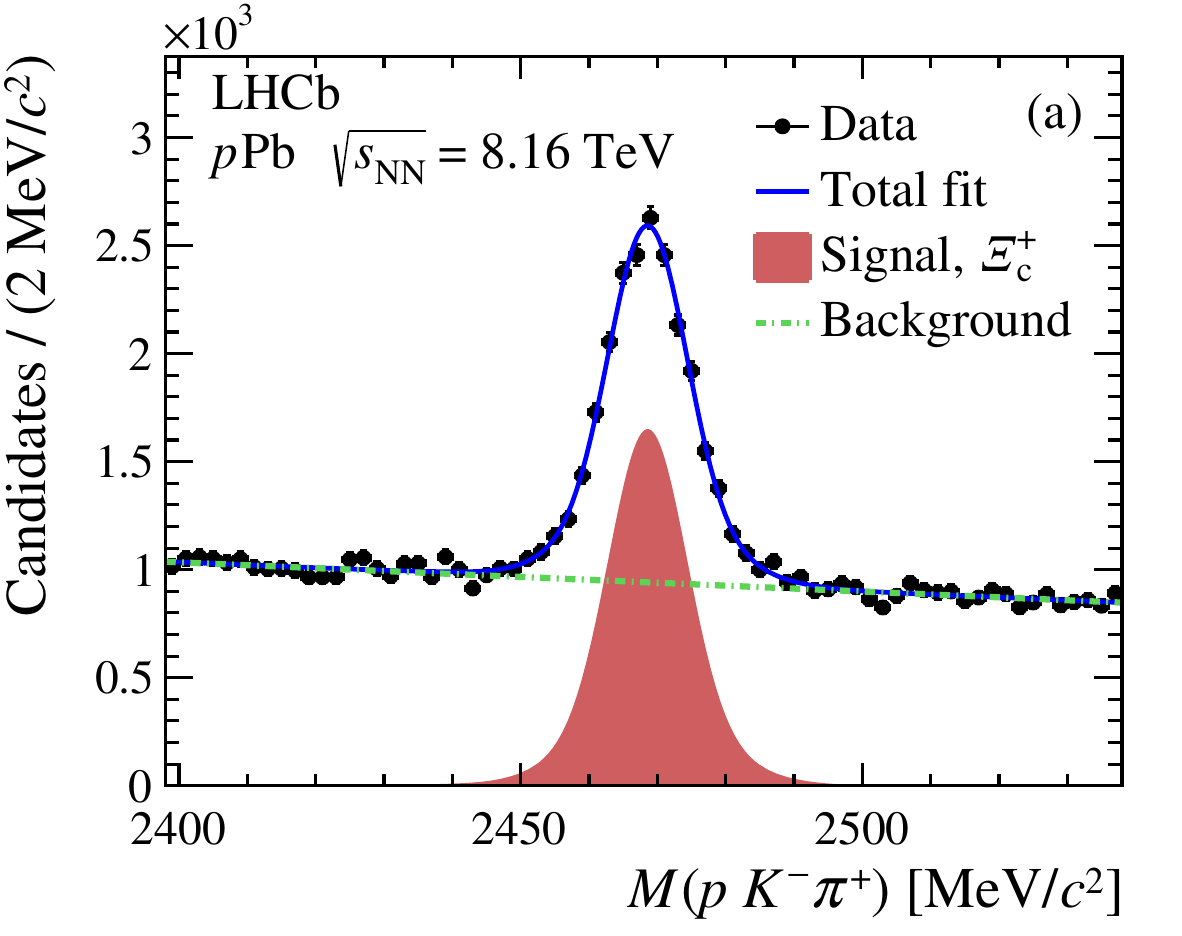}
    \includegraphics[width=0.46\linewidth]{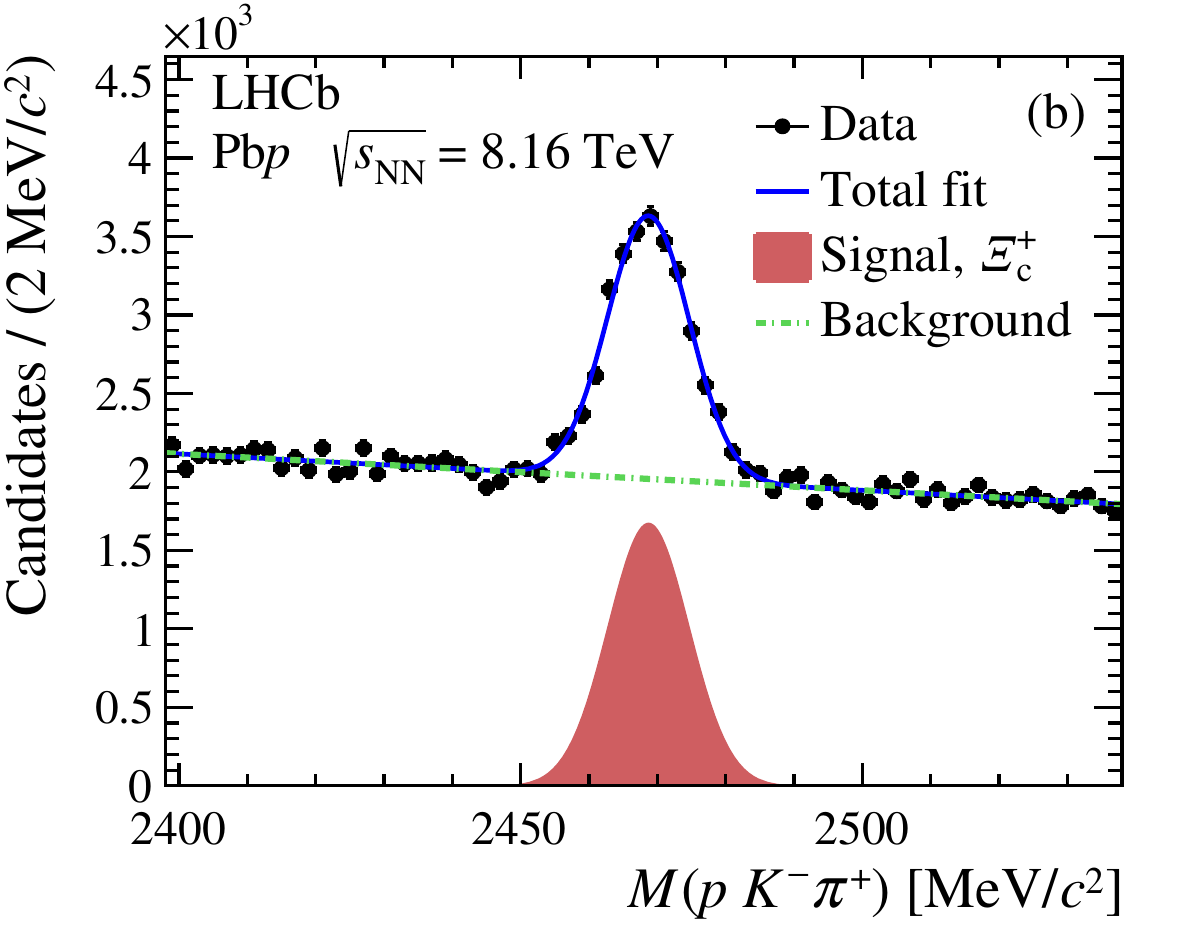}
    \includegraphics[width=0.46\linewidth]{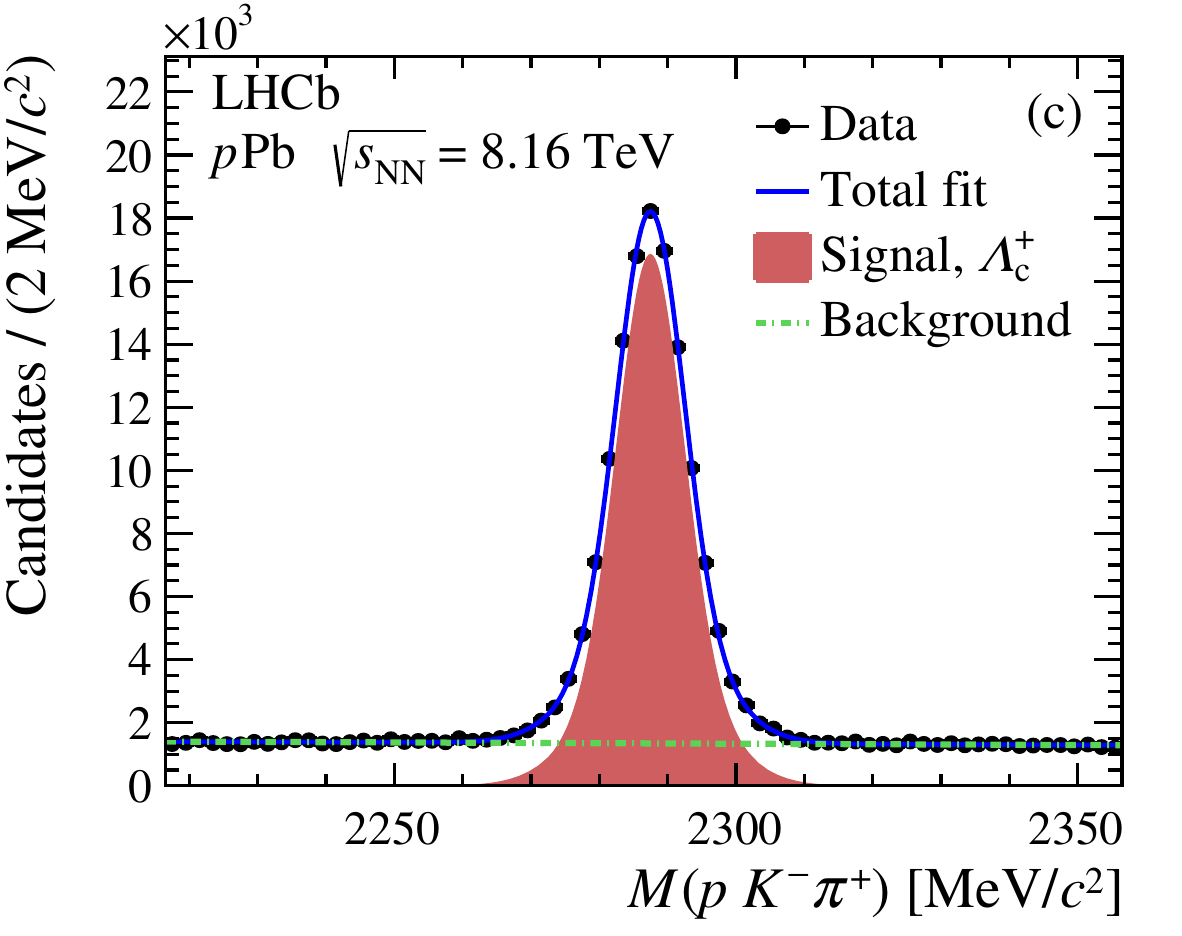}
    \includegraphics[width=0.46\linewidth]{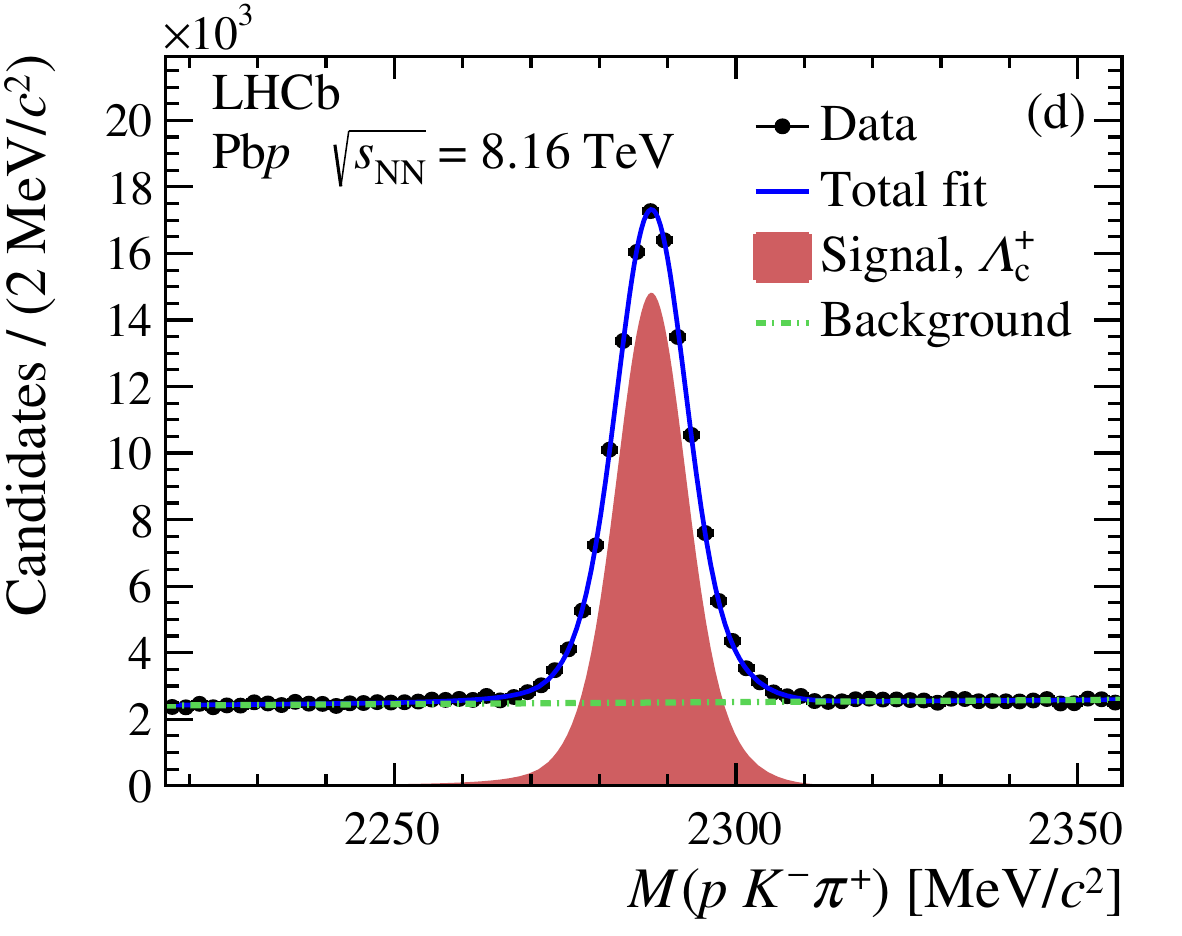}
    \vspace*{-0.5cm}
  \end{center}
   \caption{
    Invariant mass distributions for (a) \Xicp candidates in $p$Pb data, (b) \Xicp candidates in Pb$p$ data,
    (c) \Lc candidates in $p$Pb data, and (d) \Lc candidates in Pb$p$ data. The results of the fit are overlaid. \\
     }
  \label{fig:minv}
\end{figure*}

The signal yields are determined using a maximum-likelihood fit to the $\proton\Km\pip$ invariant mass distributions, which were verified to be independent of the other kinematic variables. The signal component of the fit model is represented by a sum of a Crystal Ball function~\cite{Skwarnicki:1986xj} and a Gaussian function, which share a common mean value. The background component is described by a linear function. The results of the invariant mass fits for \Xicp and \Lc in $p$Pb and Pb$p$ data samples are shown in Fig.~\ref{fig:minv}.

\begin{figure*}[t!]
  \begin{center}
    \includegraphics[width=0.49\linewidth]{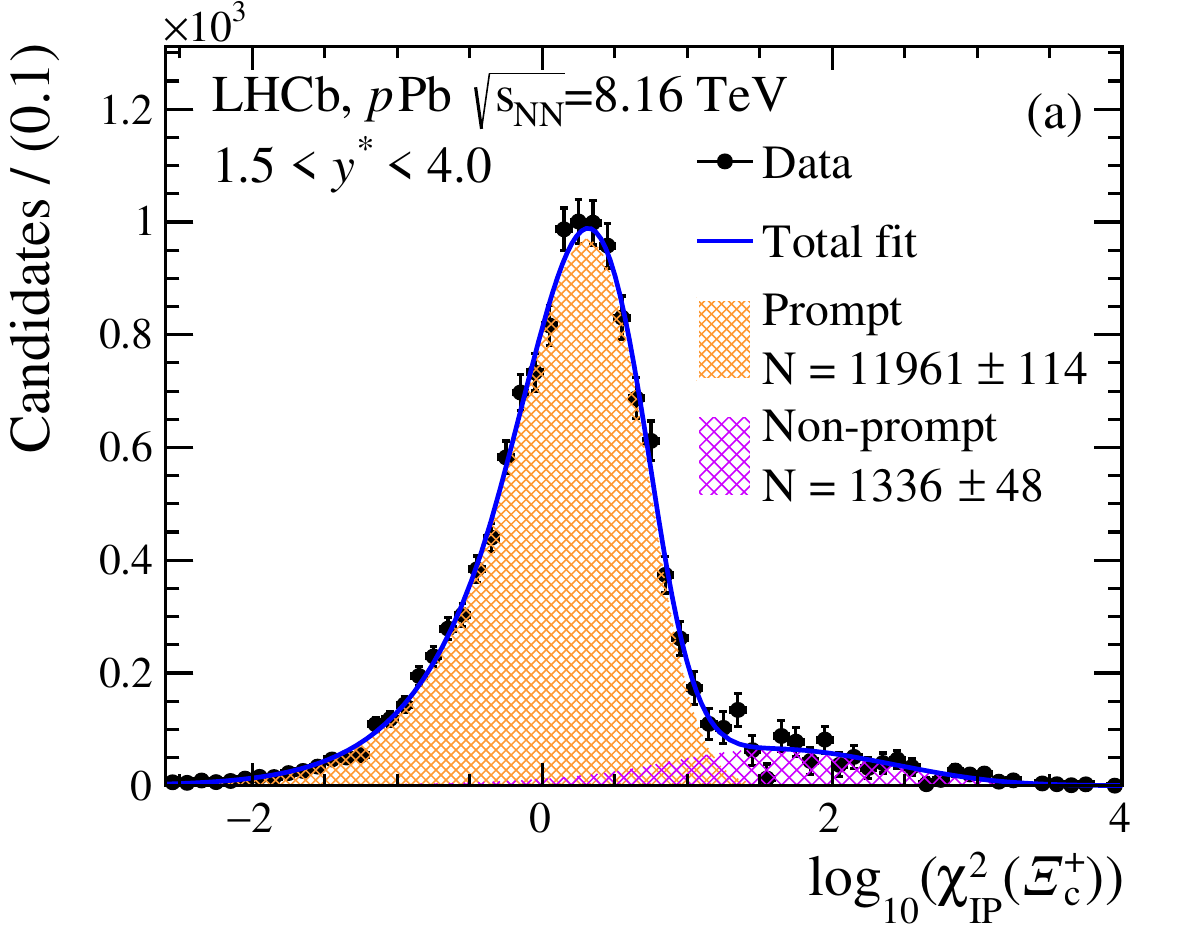}
    \includegraphics[width=0.49\linewidth]{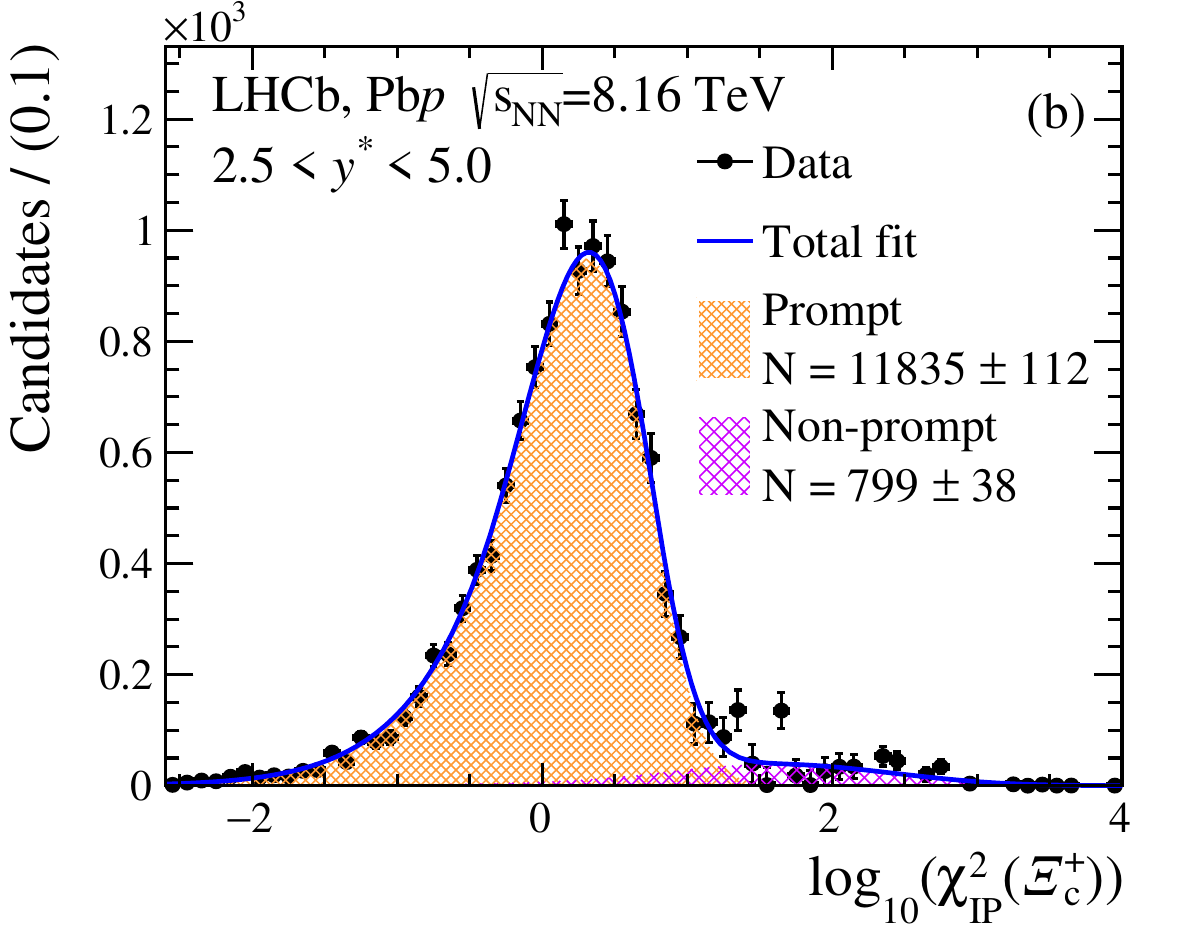}
    \includegraphics[width=0.49\linewidth]{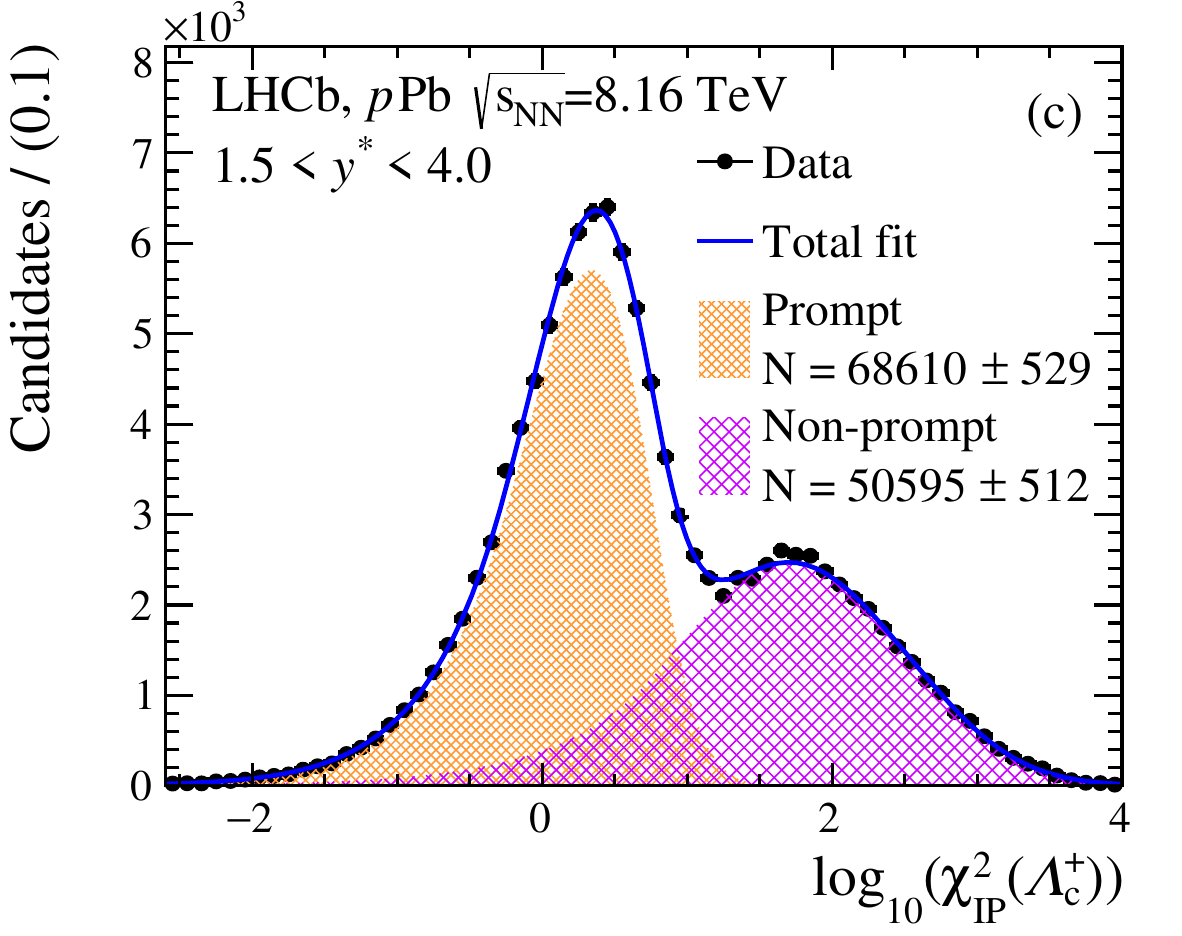}
    \includegraphics[width=0.49\linewidth]{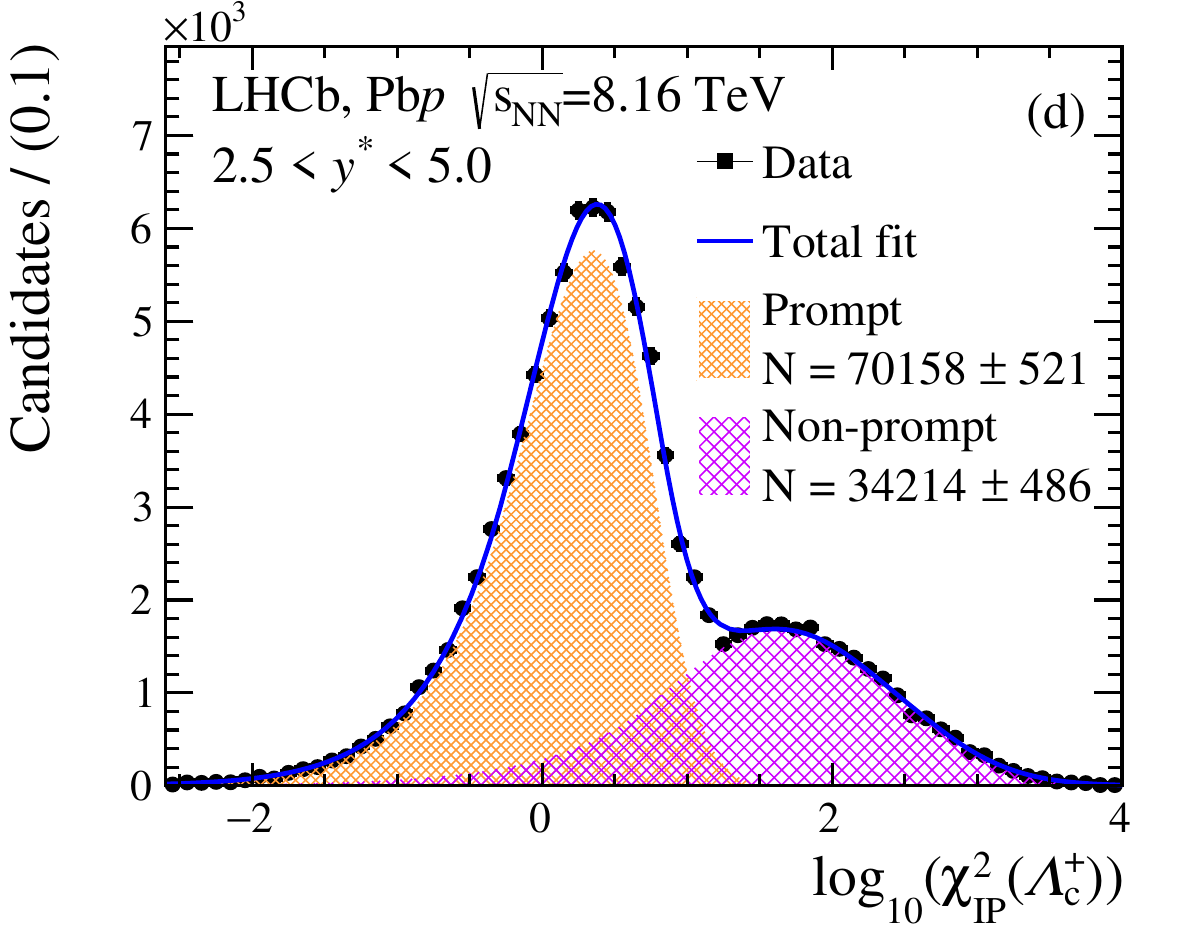}
    \vspace*{-0.5cm}
  \end{center}
  \caption{
    Background-subtracted $\log_{10}(\chisqip)$ distributions of (a) \Xicp and (b) \Lc candidates in (left) $p$Pb and (right) Pb$p$ data. The results of the fit are overlaid.}
  \label{fig:ipchi2}
\end{figure*}

The obtained signal yields are about 13.3 $\times$ $10^3$ (12.6 $\times$ $10^3$) \Xicp decays and 119.2 $\times$ $10^3$ (104.4 $\times$ $10^3$) \Lc decays in the $p$Pb (Pb$p$) sample after background subtraction, achieved using the \sPlot technique~\cite{Pivk:2004ty} with the $\proton\Km\pip$ invariant mass as the discriminating variable. The extracted signal contains promptly produced baryons and nonprompt signal from the decay of $b$ hadrons. To extract the prompt component, the method developed in Ref.~\cite{LHCb-PAPER-2018-021} is used, with the variable $\chisqip$ discriminating between prompt and nonprompt production. The $\chisqip$ variable is defined as the difference in the vertex-fit $\chi^2$ of a given PV reconstructed with and without the $\Xicp (\Lc)$ candidate under consideration. The prompt and nonprompt components are modelled with a Bukin function~\cite{BukinF}. The asymmetry parameters of the Bukin function describing the prompt component are fixed using results from fits to simulated samples. The results of the background-subtracted distribution fits of the $\log_{10}(\chisqip)$ for \Xicp and \Lc in $p$Pb and Pb$p$ data samples are shown in Fig.~\ref{fig:ipchi2}. The quantity $N^{\Xicp (\Lc)}(\pt, y^{*})$ is obtained by performing the invariant mass and $\log_{10}(\chisqip)$ fits in each (\pt, $y^*$) bin.

Simulated data are used to determine the total efficiency, defined as the product of the geometrical acceptance of the detector, and the reconstruction, selection, PID and trigger efficiencies. The samples are weighted to match the background-subtracted data using the distributions of \pt, $y^*$, the number of VELO clusters and the invariant mass of the $pK^{-}$ and $K^{-}\pi^{+}$ combinations. 
The PID efficiency is also measured with dedicated calibration data samples~\cite{LHCb-PUB-2016-021}.


\section{Systematic uncertainties}
Several sources of systematic uncertainty are studied. A systematic uncertainty in the signal yield determination is evaluated by changing the function used in the fit for the nonprompt component, switching from a Bukin function to a Gaussian. The difference in the signal yields obtained with the two fits is taken as a systematic uncertainty. The systematic uncertainty associated to the background determination, which is dominant, is evaluated by using a sideband subtraction technique, instead of the \sPlot method. The uncertainties in reconstruction, selection, and PID efficiencies are evaluated by varying the reweighting procedure by excluding the distributions of the invariant mass of the $pK^{-}$ and $K^{-}\pi^{+}$ combinations. The uncertainties in the branching fractions are accounted for. Summaries of the systematic uncertainties in the \Xicp and \Lc cross sections can be found in Tables~\ref{tab:systematicpt},~\ref{tab:systematicy}, and~\ref{tab:systema_cor}.

\begin{table*}[t!]
  \caption{
{    \small{ 
    Summary of systematic uncertainties on the \Xicp and \Lc cross sections in \pt bins in the $p$Pb and Pb$p$ samples. The range of uncertainties correspond to the values obtained for different bins of \pt.}}}
\begin{center}\begin{tabular}{lcccc}
    \hline
                        & \multicolumn{2}{c}{\Xicp} & \multicolumn{2}{c}{\Lc} \\
    $ $                 & $p$Pb &  Pb$p$ & $p$Pb & Pb$p$  \\
    \hline
    Signal              & 0.1--2.2 \%  & 0.2--2.3 \%  &     --        &     --       \\
    Background          & 1.3--5.7 \%  & 0.5--18.0 \% & 0.1--1.0 \%  & 0.1--0.8 \% \\
    \eacc               & 0.1--0.2 \%  & 0.1--0.3 \%  & 0.1--0.2 \%  & 0.1--0.2 \%  \\
    \esel               & 1.1--3.5 \%  & 1.3--4.8 \%  & 3.6--7.3 \%  & 2.7--5.5 \%  \\
    \epid               & 0.3--0.7 \%  & 0.6--1.4 \%  & 0.2--0.6 \%  & 0.5--1.1 \%  \\
    \etrg               & 0.4--0.5 \%  & 0.4--0.5 \%  & 0.1--0.6 \%  & 0.4--0.8 \%  \\
    \hline
    Total               & 2.0--6.3 \% & 2.9--18.0 \%  & 3.6--7.3 \% & 2.8--5.6 \%  \\
    \hline
\end{tabular}\end{center}
\label{tab:systematicpt}
\end{table*}

\begin{table*}[t!]
  \caption{
{    \small{ 
    Summary of systematic uncertainties on the \Xicp and \Lc cross sections in $y^*$ bins in the $p$Pb and Pb$p$ samples. The range of uncertainties correspond to the values obtained for different bins of $y^*$.}}}
\begin{center}\begin{tabular}{lcccc}
    \hline
                        & \multicolumn{2}{c}{\Xicp} & \multicolumn{2}{c}{\Lc} \\
    $ $                 & $p$Pb & Pb$p$ & $p$Pb & Pb$p$ \\
    \hline
    Signal              & 0.2--3.0 \%  & 0.2--3.6 \%  & 2.0--5.9 \%  &    --       \\
    Background          & 0.1--5.7 \%  & 1.7--27.4 \% & 0.1--4.6 \%  & 0.7--17.7 \% \\
    \eacc               & 0.1--0.4 \%  & 0.1--0.8 \%  & 0.1--0.3 \%  & 0.1--0.5 \%  \\
    \esel               & 0.7--2.8 \%  & 1.5--4.2 \%  & 3.4--6.8 \%  & 1.2--14.4 \% \\
    \epid               & 0.4--1.5 \%  & 0.5--3.0 \%  & 0.2--2.3 \%  & 0.4--3.8 \%  \\
    \etrg               & 0.4--0.5 \%  & 0.4--0.6 \%  & 0.4--0.5 \%  & 0.4--1.2 \%  \\
    \hline
    Total               & 1.6--6.4 \% & 2.7--27.8 \%  & 4.1--9.9 \% & 1.8--17.9 \%  \\
    \hline
\end{tabular}\end{center}
\label{tab:systematicy}
\end{table*}

\begin{table*}[t!]
  \caption{
{    \small{ 
    Summary of systematic uncertainties correlated among bins on the \Xicp and \Lc cross sections in the $p$Pb and Pb$p$ samples.}}}
\begin{center}\begin{tabular}{lcc}
    \hline
    $ $                 & $p$Pb & Pb$p$    \\
    \hline
    Luminosity          & 2.6 \% & 2.5 \% \\
    Signal              & 4.8 \% & 2.8--3.1 \% \\
    Tracking   & \multicolumn{2}{c}{5.5$\%$} \\
    $\mathcal{B}(\Xicp \rightarrow p^{+} K^{-} \pi^{+})$ & \multicolumn{2}{c}{48.4$\%$} \\
    $\mathcal{B}(\Lc \rightarrow p^{+} K^{-} \pi^{+})$   & \multicolumn{2}{c}{5.1$\%$} \\
    \hline

  \end{tabular}\end{center}
\label{tab:systema_cor}
\end{table*}
\section{Results}
In this section the different measurements performed are discussed. More results can be found in the Appendix.

\subsection{Double-differential cross section}
\begin{figure*}[htbp!]
  \begin{center}
      \includegraphics[width=0.9\linewidth]{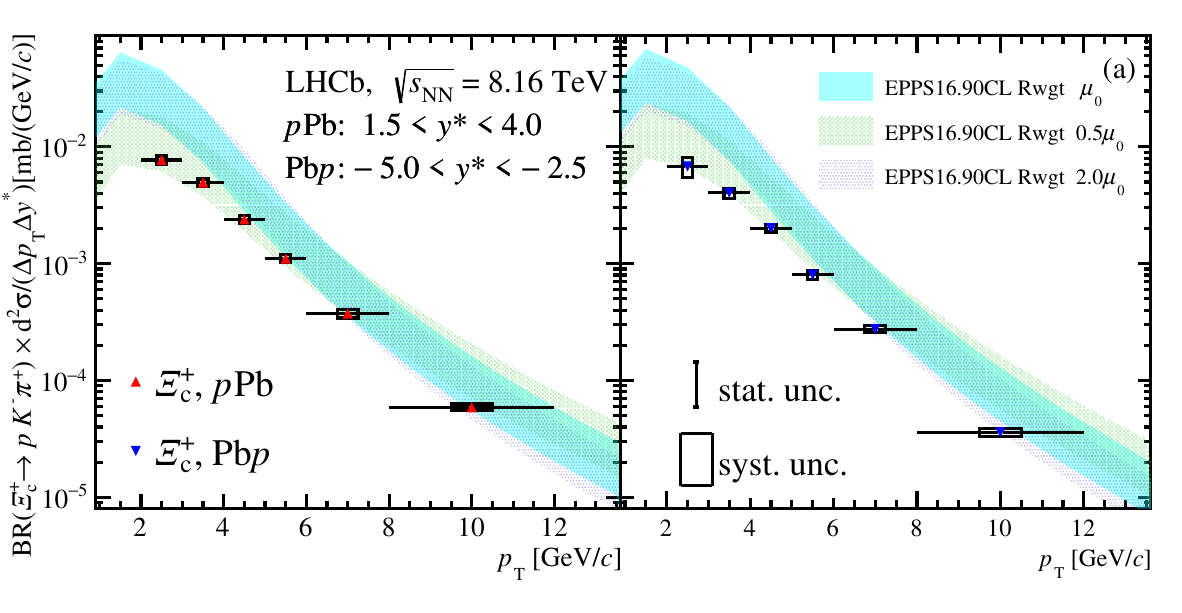}
      \includegraphics[width=0.9\linewidth]{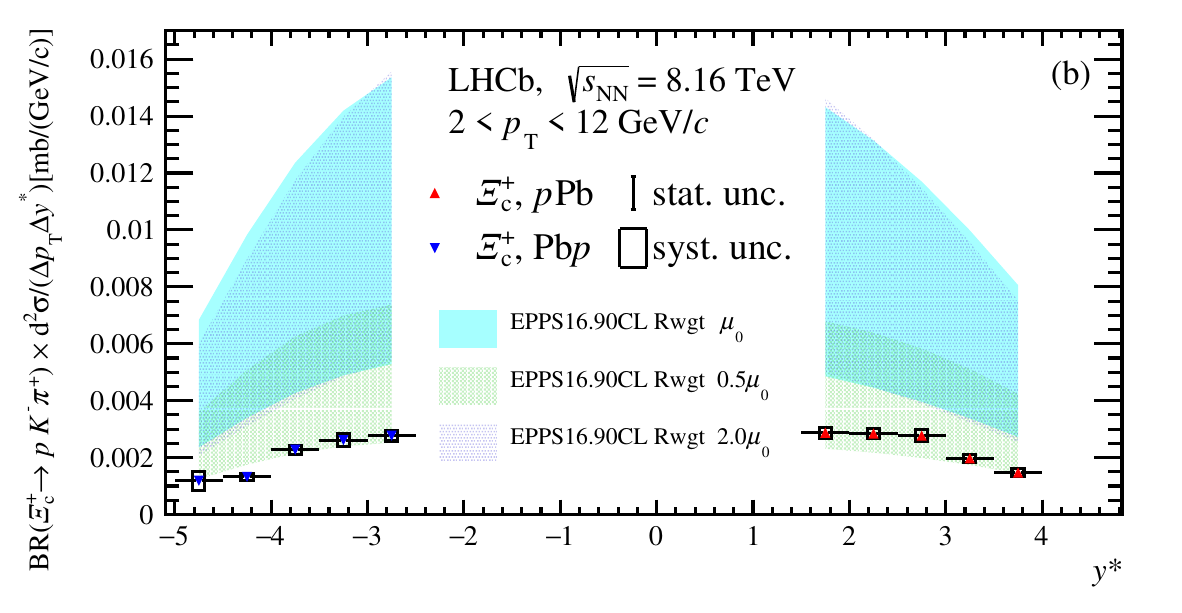}
    \vspace*{-0.5cm}
  \end{center}
  \caption{
    Double-differential cross section of the prompt \Xicp baryon times $\mathcal{B}(\Xicp \rightarrow \proton\Km\pip)$ in proton-lead collisions as a function of (a) \pt and (b) $y^*$ in $p$Pb (red triangles) and Pb$p$ (blue triangles) collisions. The error bars represent the statistical uncertainties, while the black squares represent the total systematic uncertainties which include correlations among bins.}
  \label{fig:xsec_plots}
\end{figure*}

The double-differential cross sections of prompt \Xicp production times \mbox{$\mathcal{B}(\Xicp \rightarrow \proton\Km\pip)$} in proton-lead collisions at $\sqsnn=8.16\tev$ are measured as a function of \pt integrated over $y^*$ in the regions \mbox{1.5 $<y^*<$ 4.0} and \mbox{$-$5.0 $<y^*<$ $-$2.5}, and as a function of $y^*$ integrated over the \pt range between 2.0 and 12.0~\gevc. The double-differential cross sections of prompt \Xicp production are shown in Fig.~\ref{fig:xsec_plots}. The data are compared with theoretical predictions~\cite{Lansberg:2016deg,Kusina:2017gkz,Kusina:2020dki} from the HELAC-Onia method~\cite{Shao:2012iz,Shao:2015vga} called EPPS16~\cite{Eskola:2016oht} with three factorization scale choices. The data agree with the predictions and appear to be best described using the scale $0.5\mu_{0}$. 
For the forward and backward regions, the integrated cross sections are 
\begin{align*}
    \sigma_{\Xicp}^{p\textrm{Pb}}(2<\pt<12 \gevc, 1.5<y^{*}< 4.0) = 9.69 \pm 0.12 \pm 0.26 \pm 4.72 \,\textrm{mb}, \\
    \sigma_{\Xicp}^{\textrm{Pb}p}(2<\pt<12 \gevc, -5.0<y^{*}<-2.5) = 8.10 \pm 0.11 \pm 0.72 \pm 3.95 \,\textrm{mb},
\end{align*}
where the first uncertainty is statistical, the second is the uncorrelated systematic uncertainty and the third is the systematic uncertainty fully correlated among bins.


\subsection{Forward-backward asymmetry}

\begin{figure*}[!htb]
  \begin{center}
    \includegraphics[width=0.7\linewidth]{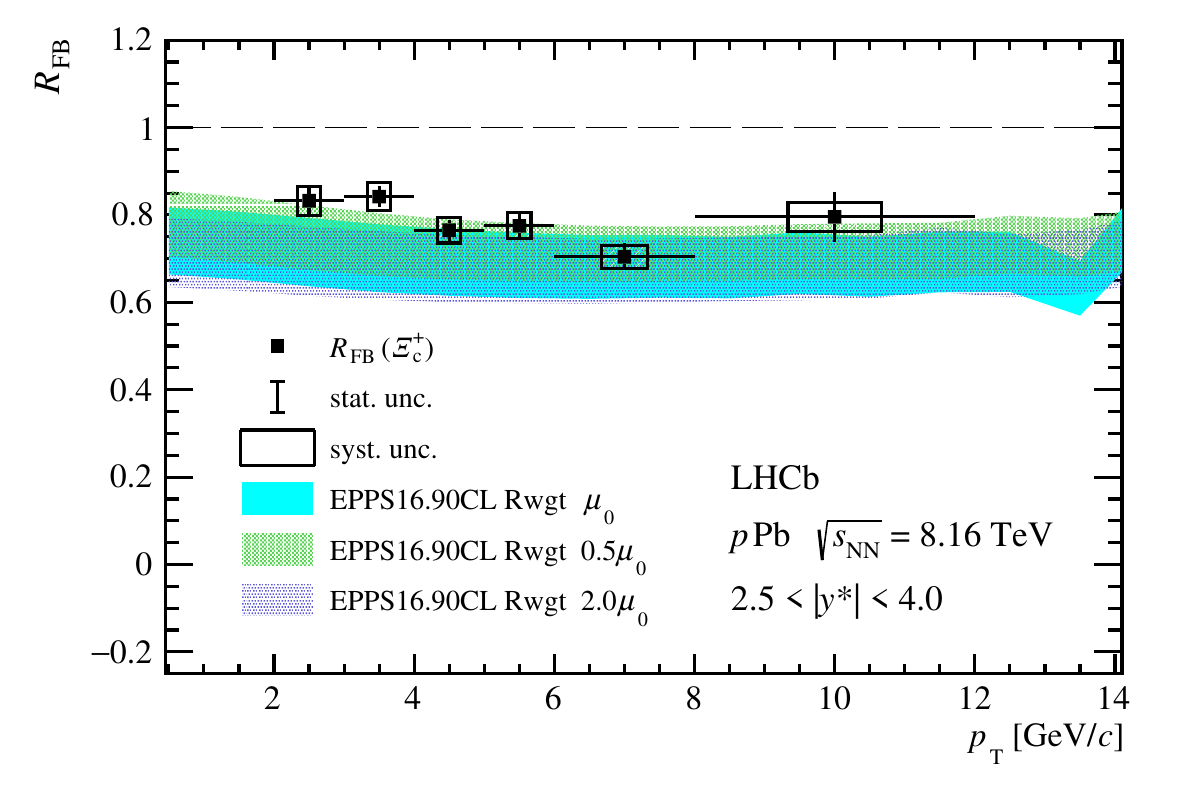}
    \vspace*{-0.5cm}
  \end{center}
  \caption{
    Forward-backward ratio of \Xicp production as a function of \pt. The error bars represent the statistical uncertainties, while the boxes indicate the systematic uncertainty.}
  \label{fig:FBratio_result}
\end{figure*}

The forward-backward ratio $R_{\textrm{FB}}$, defined in Eq. (\ref{eq:3}) and measured as a function of \pt, is shown in Fig.~\ref{fig:FBratio_result}.
The ratio is independent of \pt and agrees with the theoretical prediction within one standard deviation.

\subsection{Differential ratio of \Xicp to \Lc($D^{0}$)}

\begin{figure*}[!htb]
  \begin{center}
    \includegraphics[width=0.66\linewidth]{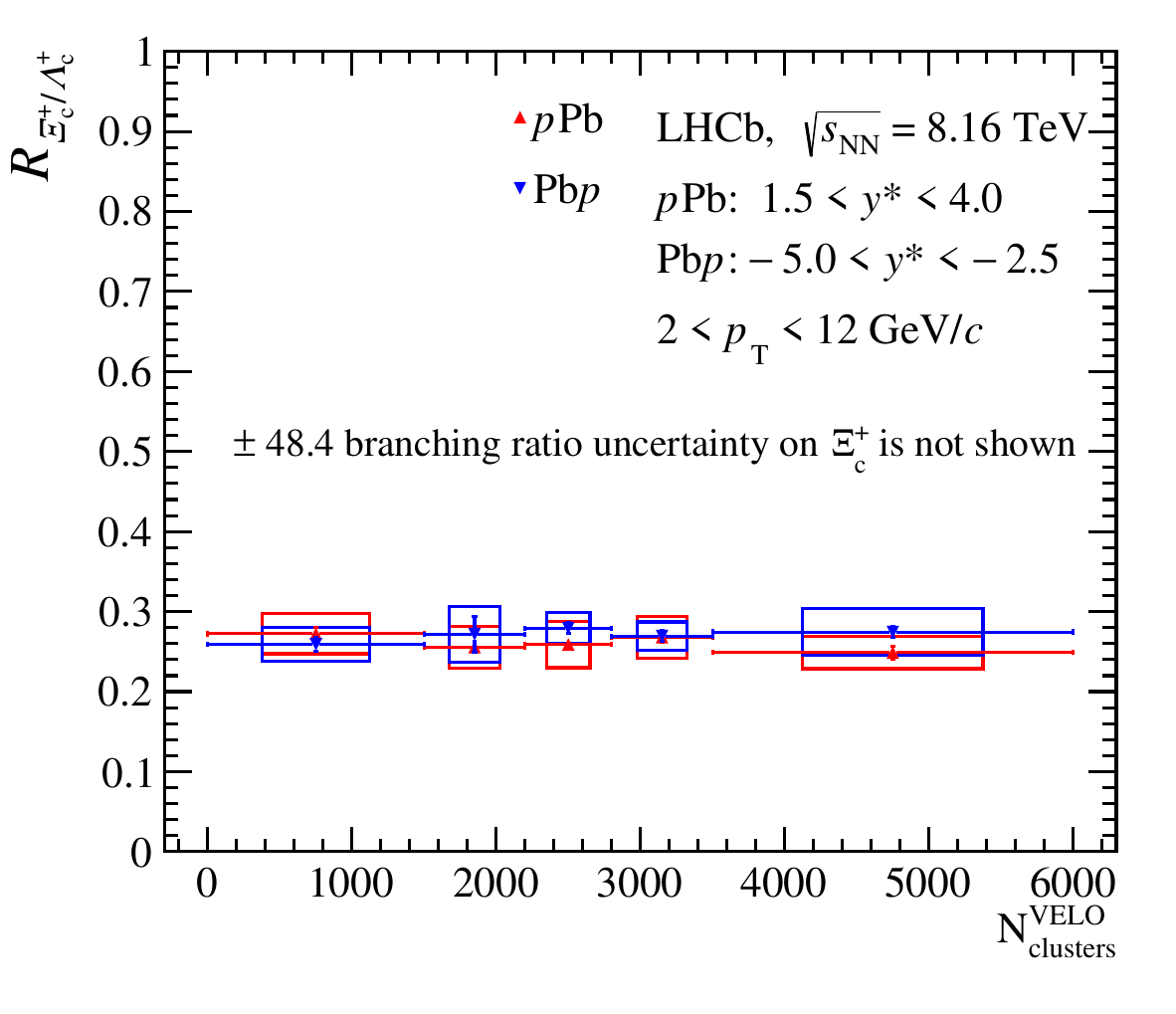}
    \vspace*{-0.5cm}
  \end{center}
  \caption{
    Ratio of prompt \Xicp to \Lc production in $p$Pb (red triangles) and Pb$p$ (blue triangles) data samples as a function of $\textrm{N}_{\textrm{clusters}}^{\textrm{VELO}}$. The error bars represent the statistical uncertainties, while the squares indicate the systematic uncertainty. The branching ratio uncertainty on \Xicp is not shown.}
  \label{fig:ratio_vc}
\end{figure*}

The differential ratio of \Xicp to \Lc production is measured as a function of the number of clusters reconstructed in the VELO ($\textrm{N}_{\textrm{clusters}}^{\textrm{VELO}}$) and given in Fig.~\ref{fig:ratio_vc}. The ratios are constant as a function of $\textrm{N}_{\textrm{clusters}}^{\textrm{VELO}}$, similarly for the $p$Pb and Pb$p$ data, and the results show no indication of strangeness enhancement.

\begin{figure*}[!htb]
  \begin{center}
    \includegraphics[width=0.46\linewidth]{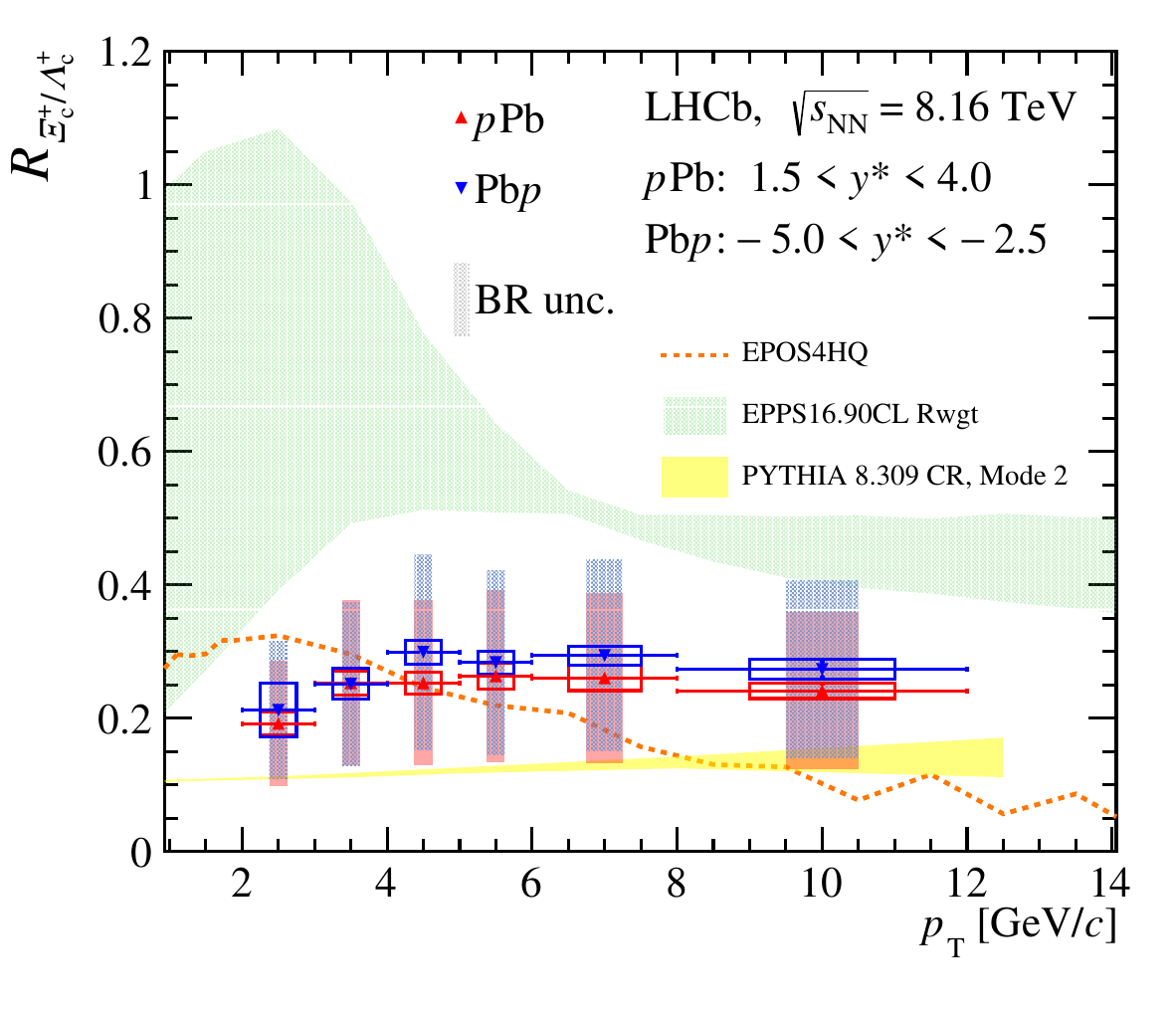}
    \includegraphics[width=0.46\linewidth]{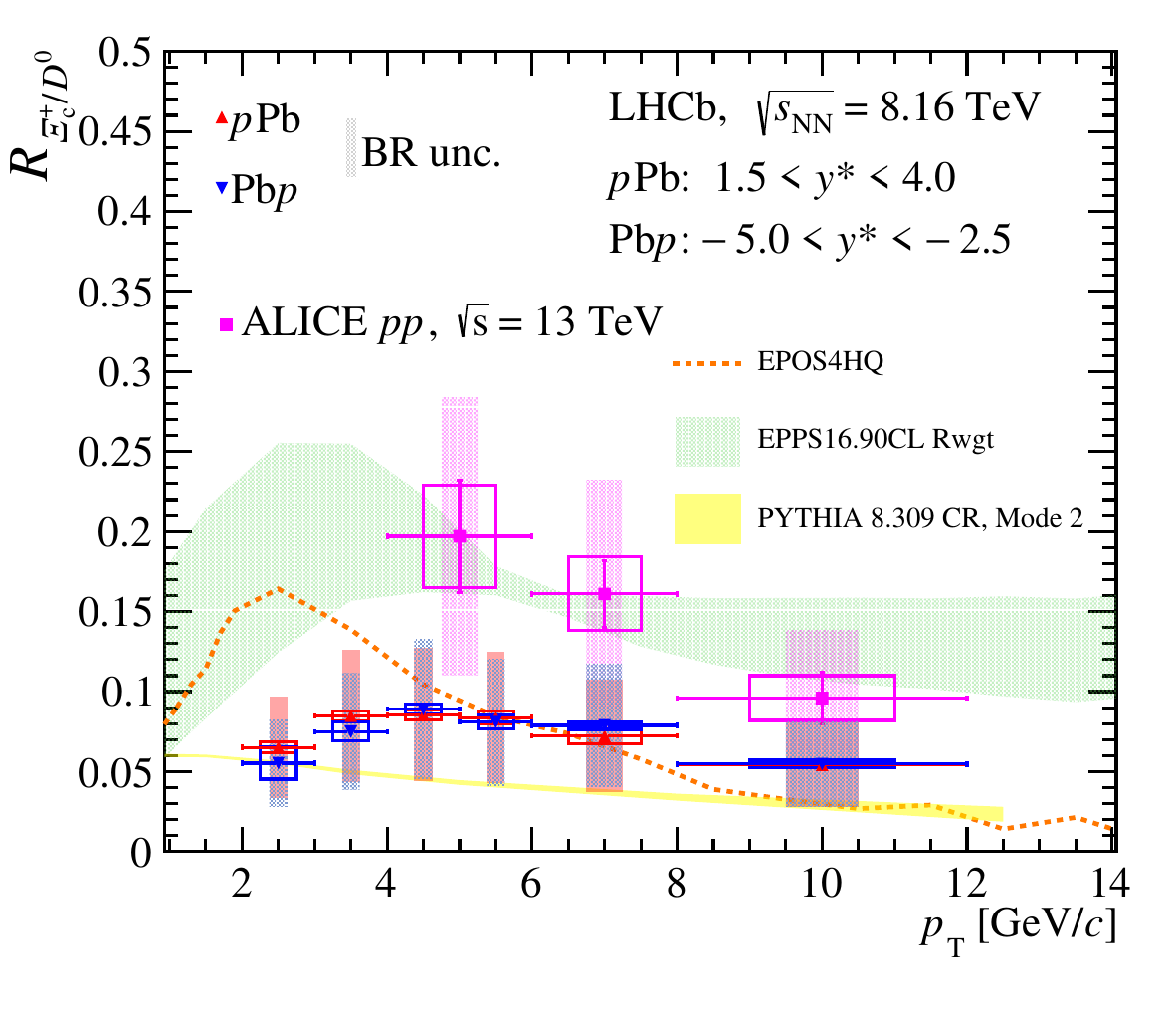}
    \vspace*{-0.5cm}
  \end{center}
  \caption{
    Ratio of (left) prompt \Xicp to \Lc production and (right) \Xicp to $D^{0}$ production in the $p$Pb (red triangles) and Pb$p$ (blue triangles) data samples as a function of \pt. The error bars represent the statistical uncertainties, while the empty rectangles indicate the systematic uncertainty. Shaded rectangles denote the branching ratio uncertainty.}
  \label{fig:ptratio_result}
\end{figure*}

Since LHCb has already measured the $D^{0}$ production cross section in $p$Pb collisions at $\sqs=8.16\tev$ \cite{LHCb-PAPER-2022-007}, it is also possible to compute the $\Xicp/D^{0}$ production ratio. The differential ratios of \Xicp to \Lc production and \Xicp to $D^{0}$ production are shown in Fig.~\ref{fig:ptratio_result} as a function of \pt.
Both the \Xicp/\Lc and \Xicp/$D^{0}$ ratios show no significant \pt dependence, similarly for the $p$Pb and Pb$p$ data samples. This result provides a strong indication that the same processes govern hadronization in $p$Pb and Pb$p$ collisions. This is the first time they are measured in this 
system.
The data are compared with EPPS16 nPDF predictions~\cite{Eskola:2016oht}, which uses the cross section measured in $pp$ collisions by ALICE~\cite{ALICE:2021bli} as input for their calculation. The EPPS16 model shows a similar trend as the data, but significantly overestimates it. The measurements are also compared with results from the \pythia 8.3 event generator with a tune that implements color reconnection (CR) beyond the leading-color approximation~\cite{Christiansen:2015yqa} and the computation obtained with EPOS4HQ, the heavy quark extension of the new EPOS4 framework~\cite{Werner:2023fne}. Both computations are based on results in $pp$ collisions. The \Xicp/\Lc cross section ratio is best described by the EPOS4HQ model within uncertainties, despite showing a different trend, especially at low \pt. This behavior is even more pronounced in the \Xicp/$D^{0}$ cross section ratio.
The \Xicp to $D^{0}$ production ratio is also compared with the result of ALICE at \sqs = 13 \tev at $|y|<$0.5 in $pp$ collisions~\cite{ALICE:2021bli}. The ALICE result is generally higher, but the two measurements agree within the uncertainties.

\section{Summary}

In summary, the prompt \Xicp production cross section in $p$Pb and Pb$p$ collisions at a center-of-mass energy of $\sqsnn=8.16$\tev at the LHCb experiment is measured differentially for the first time as a function of \pt and $y^{*}$. The cross section measurement provides new constraints for nPDF calculations, especially at low \pt, where the uncertainty on the factorisation scale is the largest. The forward-backward ratio $R_{\textrm{FB}}$ is measured and found to be well described by nuclear shadowing calculations showing that there are no major final-state effects involved, in contrast to what has been observed in $D^0$ production studies in the same kinematic range at the same energy~\cite{LHCb-PAPER-2022-007}. Prompt \Xicp production is compared to prompt \Lc production in the same kinematic region and their ratio is found to be constant within the uncertainties as a function of \pt and multiplicity, which is an indication that similar effects govern both \Xicp and \Lc production. The ratio of \Xicp to $D^{0}$ production is also measured as a function of \pt. Both ratios are found to be similar in $p$Pb and Pb$p$ collisions but they show different trends compared to $pp$ theory calculations. Therefore, our results show that hadronization in $p$Pb is not well understood and provides clear input for the hadronization studies in $p$Pb collisions.


\section*{Acknowledgements}
%
%
\noindent We would like to thank to Hua-Sheng Shao, Jiaxing Zhao and Joerg Aichelin for providing LHCb-specific theoretical predictions. 
We express our gratitude to our colleagues in the CERN
accelerator departments for the excellent performance of the LHC. We
thank the technical and administrative staff at the LHCb
institutes.
We acknowledge support from CERN and from the national agencies:
CAPES, CNPq, FAPERJ and FINEP (Brazil); 
MOST and NSFC (China); 
CNRS/IN2P3 (France); 
BMBF, DFG and MPG (Germany); 
INFN (Italy); 
NWO (Netherlands); 
MNiSW and NCN (Poland); 
MCID/IFA (Romania); 
MICINN (Spain); 
SNSF and SER (Switzerland); 
NASU (Ukraine); 
STFC (United Kingdom); 
DOE NP and NSF (USA).
We acknowledge the computing resources that are provided by CERN, IN2P3
(France), KIT and DESY (Germany), INFN (Italy), SURF (Netherlands),
PIC (Spain), GridPP (United Kingdom), 
CSCS (Switzerland), IFIN-HH (Romania), CBPF (Brazil),
and Polish WLCG (Poland).
We are indebted to the communities behind the multiple open-source
software packages on which we depend.
Individual groups or members have received support from
ARC and ARDC (Australia);
Key Research Program of Frontier Sciences of CAS, CAS PIFI, CAS CCEPP, 
Fundamental Research Funds for the Central Universities, 
and Sci. \& Tech. Program of Guangzhou (China);
Minciencias (Colombia);
EPLANET, Marie Sk\l{}odowska-Curie Actions, ERC and NextGenerationEU (European Union);
A*MIDEX, ANR, IPhU and Labex P2IO, and R\'{e}gion Auvergne-Rh\^{o}ne-Alpes (France);
AvH Foundation (Germany);
ICSC (Italy); 
GVA, XuntaGal, GENCAT, Inditex, InTalent and Prog.~Atracci\'on Talento, CM (Spain);
SRC (Sweden);
the Leverhulme Trust, the Royal Society
 and UKRI (United Kingdom).



\appendix
\section{Additional plots}
\label{sec:appendix}

\begin{figure*}[!htb]
  \begin{center}
    \includegraphics[width=0.70\linewidth]{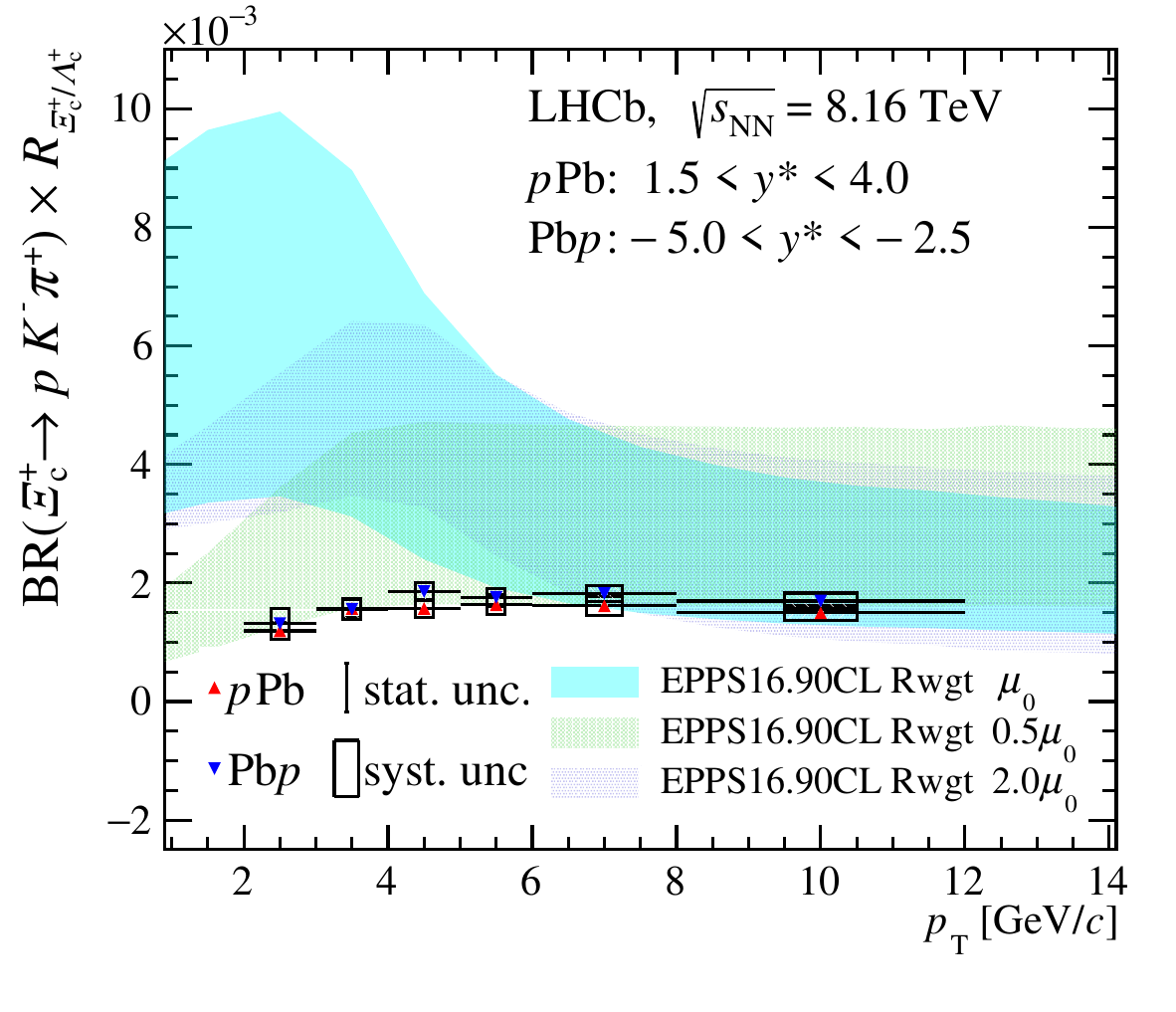}
    \vspace*{-0.5cm}
  \end{center}
  \caption{
    Production ratio of prompt \Xicp to \Lc baryons multiplied by $\mathcal{B}(\Xicp \rightarrow \proton\Km\pip)$ in $p$Pb (red triangles) and Pb$p$ (blue triangles) data samples as a function of \pt. The error bars represent the statistical uncertainties while the squares indicate the systematic uncertainty.}
  \label{fig:ratio_xiclc_br_pt}
\end{figure*}

\begin{figure*}[!htb]
  \begin{center}
    \includegraphics[width=0.70\linewidth]{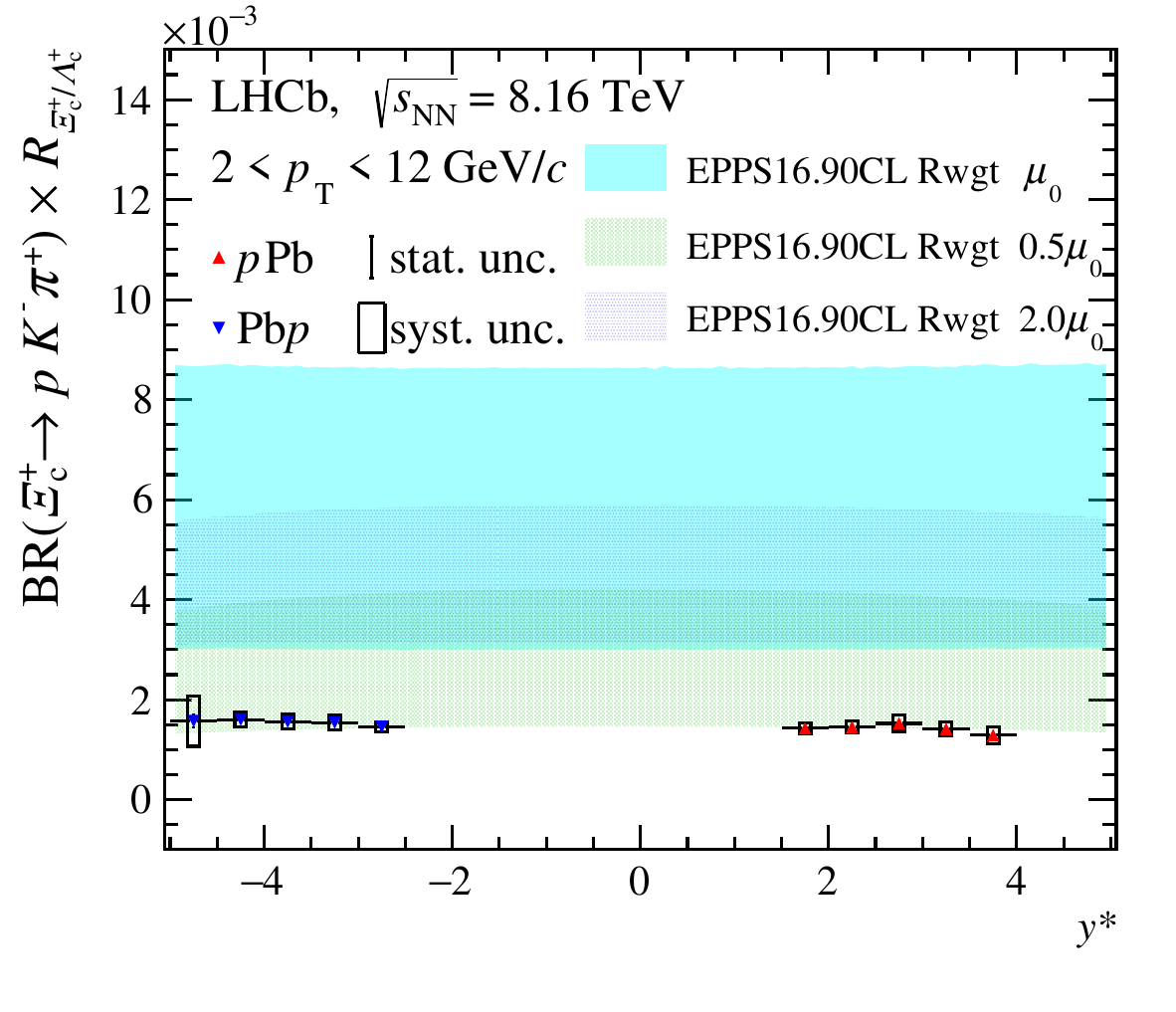}
    \vspace*{-0.5cm}
  \end{center}
  \caption{
    Production ratio of prompt \Xicp to \Lc baryons multiplied by $\mathcal{B}(\Xicp \rightarrow \proton\Km\pip)$ in $p$Pb (red triangles) and Pb$p$ (blue triangles) data samples as a function of $y^{*}$. The error bars represent the statistical uncertainties while the squares indicate the systematic uncertainty.}
  \label{fig:ratio_xiclc_br_y}
\end{figure*}

\begin{figure*}[h]
  \begin{center}
    \includegraphics[width=0.70\linewidth]{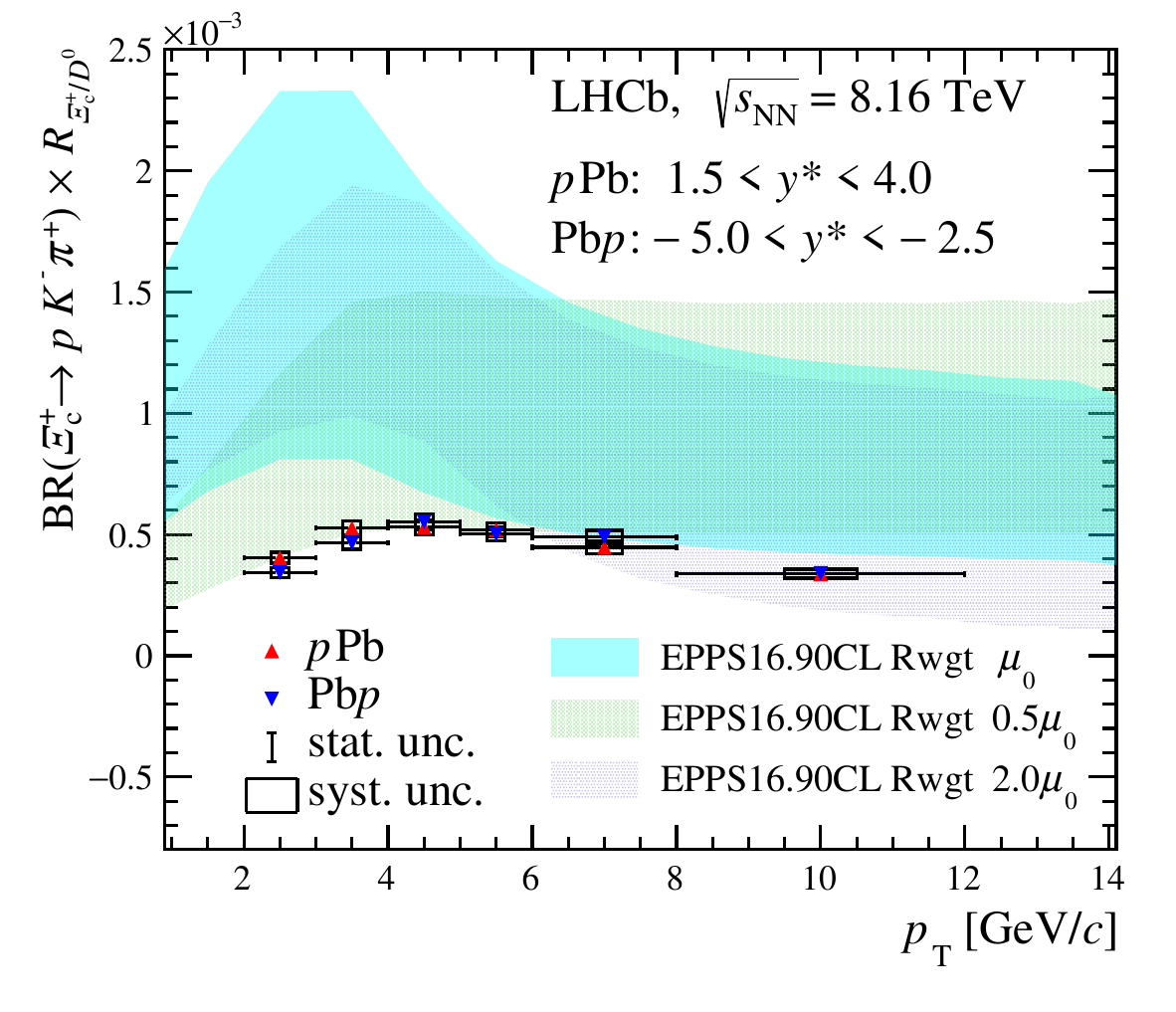}
    \vspace*{-0.5cm}
  \end{center}
  \caption{
    Production ratio of prompt \Xicp to $D^{0}$ baryons multiplied by $\mathcal{B}(\Xicp \rightarrow \proton\Km\pip)$ in $p$Pb (red triangles) and Pb$p$ (blue triangles) data samples as a function of \pt. The error bars represent the statistical uncertainties while the squares indicate the systematic uncertainty.}
  \label{fig:ratio_xicd0_br_pt}
\end{figure*}

\begin{figure*}[h]
  \begin{center}
    \includegraphics[width=0.70\linewidth]{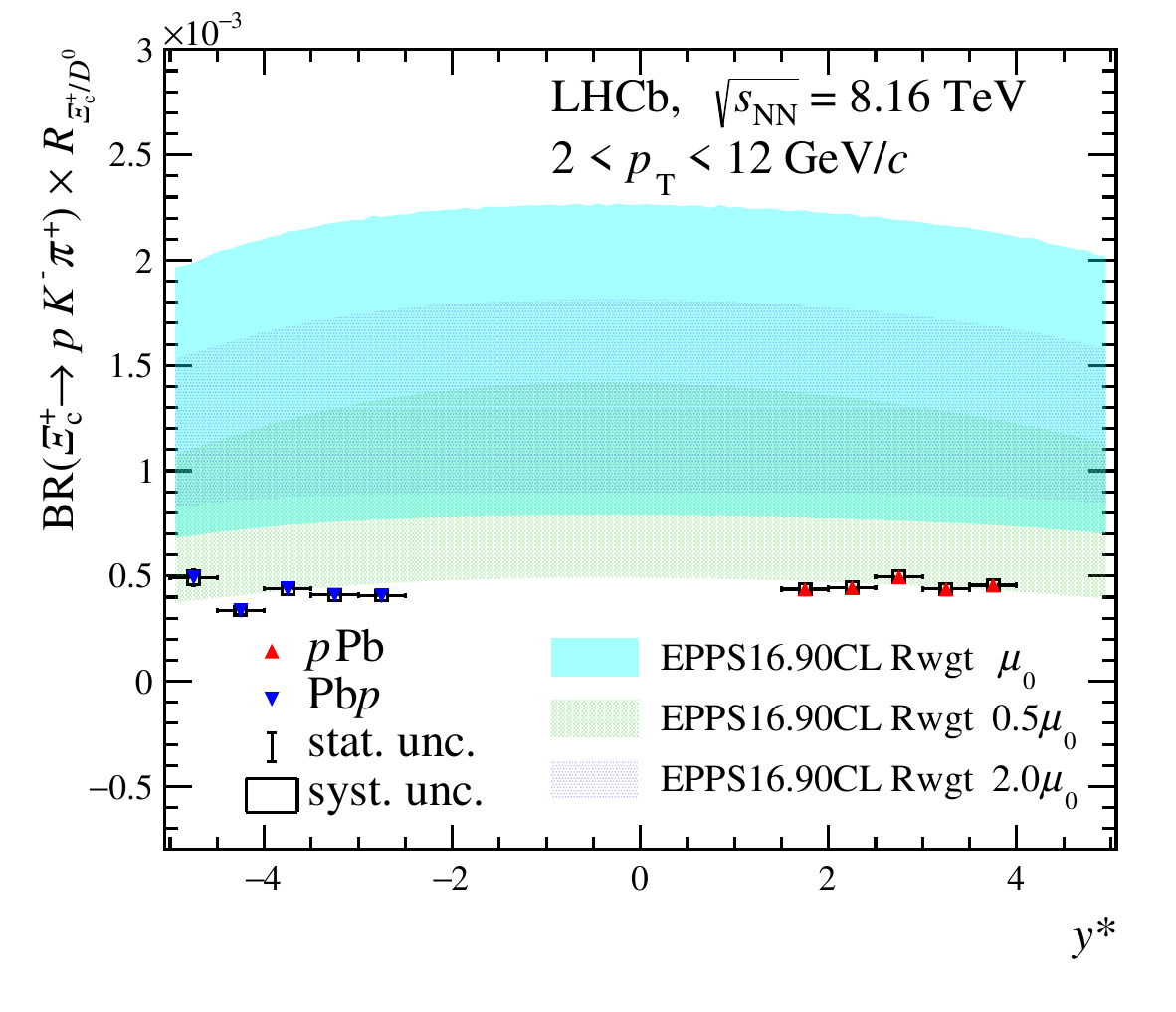}
    \vspace*{-0.5cm}
  \end{center}
  \caption{
    Production ratio of prompt \Xicp to $D^{0}$ baryons multiplied by $\mathcal{B}(\Xicp \rightarrow \proton\Km\pip)$ in $p$Pb (red triangles) and Pb$p$ (blue triangles) data samples as a function of $y^{*}$. The error bars represent the statistical uncertainties while the squares indicate the systematic uncertainty.}
  \label{fig:ratio_xicd0_br_y}
\end{figure*}

 The differential ratios of \Xicp to \Lc and to $D^{0}$ production multiplied by $\mathcal{B}(\Xicp \rightarrow \proton\Km\pip)$ are shown in Figs.~\ref{fig:ratio_xiclc_br_pt},~\ref{fig:ratio_xiclc_br_y},~\ref{fig:ratio_xicd0_br_pt} and~\ref{fig:ratio_xicd0_br_y} as a function of \pt and $y^{*}$.

\clearpage

\addcontentsline{toc}{section}{References}
\bibliographystyle{LHCb}
\bibliography{main,standard,LHCb-PAPER,LHCb-CONF,LHCb-DP,LHCb-TDR}

\ifx\mcitethebibliography\mciteundefinedmacro
\PackageError{LHCb.bst}{mciteplus.sty has not been loaded}
{This bibstyle requires the use of the mciteplus package.}\fi
\providecommand{\href}[2]{#2}
\begin{mcitethebibliography}{10}
\mciteSetBstSublistMode{n}
\mciteSetBstMaxWidthForm{subitem}{\alph{mcitesubitemcount})}
\mciteSetBstSublistLabelBeginEnd{\mcitemaxwidthsubitemform\space}
{\relax}{\relax}

\bibitem{LHCb-PAPER-2016-042}
LHCb collaboration, R.~Aaij {\em et~al.},
  \ifthenelse{\boolean{articletitles}}{\emph{{Measurements of prompt charm
  production cross-sections in \proton\proton collisions at $\sqs = $5\tev}},
  }{}\href{https://doi.org/10.1007/JHEP06(2017)147}{JHEP \textbf{06} (2017)
  147}, \href{http://arxiv.org/abs/1610.02230}{{\normalfont\ttfamily
  arXiv:1610.02230}}\relax
\mciteBstWouldAddEndPuncttrue
\mciteSetBstMidEndSepPunct{\mcitedefaultmidpunct}
{\mcitedefaultendpunct}{\mcitedefaultseppunct}\relax
\EndOfBibitem
\bibitem{Collins:1985gm}
J.~C. Collins, D.~E. Soper, and G.~F. Sterman,
  \ifthenelse{\boolean{articletitles}}{\emph{{Heavy particle production in
  high-energy hadron collisions}},
  }{}\href{https://doi.org/10.1016/0550-3213(86)90026-X}{Nucl.\ Phys.\
  \textbf{B263} (1986) 37}\relax
\mciteBstWouldAddEndPuncttrue
\mciteSetBstMidEndSepPunct{\mcitedefaultmidpunct}
{\mcitedefaultendpunct}{\mcitedefaultseppunct}\relax
\EndOfBibitem
\bibitem{LHCb-PAPER-2011-018}
LHCb collaboration, R.~Aaij {\em et~al.},
  \ifthenelse{\boolean{articletitles}}{\emph{{Measurement of \bquark hadron
  production fractions in 7\tev\ \proton\proton collisions}},
  }{}\href{https://doi.org/10.1103/PhysRevD.85.032008}{Phys.\ Rev.\
  \textbf{D85} (2012) 032008},
  \href{http://arxiv.org/abs/1111.2357}{{\normalfont\ttfamily
  arXiv:1111.2357}}\relax
\mciteBstWouldAddEndPuncttrue
\mciteSetBstMidEndSepPunct{\mcitedefaultmidpunct}
{\mcitedefaultendpunct}{\mcitedefaultseppunct}\relax
\EndOfBibitem
\bibitem{Werner:2023fne}
K.~Werner and B.~Guiot,
  \ifthenelse{\boolean{articletitles}}{\emph{{Perturbative QCD concerning light
  and heavy flavor in the EPOS4 framework}},
  }{}\href{https://doi.org/10.1103/PhysRevC.108.034904}{Phys.\ Rev.\
  \textbf{C108} (2023) 034904},
  \href{http://arxiv.org/abs/2306.02396}{{\normalfont\ttfamily
  arXiv:2306.02396}}\relax
\mciteBstWouldAddEndPuncttrue
\mciteSetBstMidEndSepPunct{\mcitedefaultmidpunct}
{\mcitedefaultendpunct}{\mcitedefaultseppunct}\relax
\EndOfBibitem
\bibitem{ALICE:2017thy}
ALICE collaboration, S.~Acharya {\em et~al.},
  \ifthenelse{\boolean{articletitles}}{\emph{{$\Lambda_{\rm c}^+$ production in
  pp collisions at $\sqrt{s} = 7$ TeV and in p-Pb collisions at $\sqrt{s_{\rm
  NN}} = 5.02$ TeV}}, }{}\href{https://doi.org/10.1007/JHEP04(2018)108}{JHEP
  \textbf{04} (2018) 108},
  \href{http://arxiv.org/abs/1712.09581}{{\normalfont\ttfamily
  arXiv:1712.09581}}\relax
\mciteBstWouldAddEndPuncttrue
\mciteSetBstMidEndSepPunct{\mcitedefaultmidpunct}
{\mcitedefaultendpunct}{\mcitedefaultseppunct}\relax
\EndOfBibitem
\bibitem{LHCb-PAPER-2018-021}
LHCb collaboration, R.~Aaij {\em et~al.},
  \ifthenelse{\boolean{articletitles}}{\emph{{Prompt \Lc production in
  \proton{}Pb collisions at $\sqsnn = 5.02$\tev}},
  }{}\href{https://doi.org/10.1007/JHEP02(2019)102}{JHEP \textbf{02} (2019)
  102}, \href{http://arxiv.org/abs/1809.01404}{{\normalfont\ttfamily
  arXiv:1809.01404}}\relax
\mciteBstWouldAddEndPuncttrue
\mciteSetBstMidEndSepPunct{\mcitedefaultmidpunct}
{\mcitedefaultendpunct}{\mcitedefaultseppunct}\relax
\EndOfBibitem
\bibitem{ARGUS:1991vjh}
ARGUS collaboration, H.~Albrecht {\em et~al.},
  \ifthenelse{\boolean{articletitles}}{\emph{{Inclusive production of $D^0$,
  $D^+$ and $D^{*+}$ (2010) mesons in B decays and nonresonant $e^+e^-$
  annihilation at $10.6$ \gev}},
  }{}\href{https://doi.org/10.1007/BF01559430}{Z.\ Phys.\  \textbf{C52} (1991)
  353}\relax
\mciteBstWouldAddEndPuncttrue
\mciteSetBstMidEndSepPunct{\mcitedefaultmidpunct}
{\mcitedefaultendpunct}{\mcitedefaultseppunct}\relax
\EndOfBibitem
\bibitem{CLEO:1990unu}
CLEO collaboration, P.~Avery {\em et~al.},
  \ifthenelse{\boolean{articletitles}}{\emph{{Inclusive production of the
  charmed baryon $\Lambda_{c}^{+}$ from $e^+e^-$ annihilations at
  $\mathbf{\sqrt{s}} = 10.55$ \gev}},
  }{}\href{https://doi.org/10.1103/PhysRevD.43.3599}{Phys.\ Rev.\  \textbf{D43}
  (1991) 3599}\relax
\mciteBstWouldAddEndPuncttrue
\mciteSetBstMidEndSepPunct{\mcitedefaultmidpunct}
{\mcitedefaultendpunct}{\mcitedefaultseppunct}\relax
\EndOfBibitem
\bibitem{ZEUS:2005pvv}
ZEUS collaboration, S.~Chekanov {\em et~al.},
  \ifthenelse{\boolean{articletitles}}{\emph{{Measurement of charm
  fragmentation ratios and fractions in photoproduction at HERA}},
  }{}\href{https://doi.org/10.1140/epjc/s2005-02397-3}{Eur.\ Phys.\ J.\
  \textbf{C44} (2005) 351},
  \href{http://arxiv.org/abs/hep-ex/0508019}{{\normalfont\ttfamily
  arXiv:hep-ex/0508019}}\relax
\mciteBstWouldAddEndPuncttrue
\mciteSetBstMidEndSepPunct{\mcitedefaultmidpunct}
{\mcitedefaultendpunct}{\mcitedefaultseppunct}\relax
\EndOfBibitem
\bibitem{ZEUS:2010cic}
ZEUS collaboration, H.~Abramowicz {\em et~al.},
  \ifthenelse{\boolean{articletitles}}{\emph{{Measurement of $D^+$ and
  $\Lambda_{c}^{+}$ production in deep inelastic scattering at HERA}},
  }{}\href{https://doi.org/10.1007/JHEP11(2010)009}{JHEP \textbf{11} (2010)
  009}, \href{http://arxiv.org/abs/1007.1945}{{\normalfont\ttfamily
  arXiv:1007.1945}}\relax
\mciteBstWouldAddEndPuncttrue
\mciteSetBstMidEndSepPunct{\mcitedefaultmidpunct}
{\mcitedefaultendpunct}{\mcitedefaultseppunct}\relax
\EndOfBibitem
\bibitem{Braaten:1994bz}
E.~Braaten, K.-m. Cheung, S.~Fleming, and T.~C. Yuan,
  \ifthenelse{\boolean{articletitles}}{\emph{{Perturbative QCD fragmentation
  functions as a model for heavy quark fragmentation}},
  }{}\href{https://doi.org/10.1103/PhysRevD.51.4819}{Phys.\ Rev.\  \textbf{D51}
  (1995) 4819},
  \href{http://arxiv.org/abs/hep-ph/9409316}{{\normalfont\ttfamily
  arXiv:hep-ph/9409316}}\relax
\mciteBstWouldAddEndPuncttrue
\mciteSetBstMidEndSepPunct{\mcitedefaultmidpunct}
{\mcitedefaultendpunct}{\mcitedefaultseppunct}\relax
\EndOfBibitem
\bibitem{ALICE:2017dja}
ALICE collaboration, S.~Acharya {\em et~al.},
  \ifthenelse{\boolean{articletitles}}{\emph{{First measurement of $\Xi_{\rm
  c}^0$ production in pp collisions at $\mathbf{\sqrt{s}} = 7$ TeV}},
  }{}\href{https://doi.org/10.1016/j.physletb.2018.03.061}{Phys.\ Lett.\
  \textbf{B781} (2018) 8},
  \href{http://arxiv.org/abs/1712.04242}{{\normalfont\ttfamily
  arXiv:1712.04242}}\relax
\mciteBstWouldAddEndPuncttrue
\mciteSetBstMidEndSepPunct{\mcitedefaultmidpunct}
{\mcitedefaultendpunct}{\mcitedefaultseppunct}\relax
\EndOfBibitem
\bibitem{ALICE:2021bli}
ALICE collaboration, S.~Acharya {\em et~al.},
  \ifthenelse{\boolean{articletitles}}{\emph{{Measurement of the cross sections
  of $\Xi^0_{c}$ and $\Xi^+_{c}$ baryons and of the branching-fraction ratio
  BR($\Xi^0_{c} \rightarrow \Xi^-{e}^+\nu_{ e}$)/BR($\Xi^0_{c} \rightarrow
  \Xi^-\pi^+$) in pp collisions at 13 TeV}},
  }{}\href{https://doi.org/10.1103/PhysRevLett.127.272001}{Phys.\ Rev.\ Lett.\
  \textbf{127} (2021) 272001},
  \href{http://arxiv.org/abs/2105.05187}{{\normalfont\ttfamily
  arXiv:2105.05187}}\relax
\mciteBstWouldAddEndPuncttrue
\mciteSetBstMidEndSepPunct{\mcitedefaultmidpunct}
{\mcitedefaultendpunct}{\mcitedefaultseppunct}\relax
\EndOfBibitem
\bibitem{Christiansen:2015yqa}
J.~R. Christiansen and P.~Z. Skands,
  \ifthenelse{\boolean{articletitles}}{\emph{{String formation beyond leading
  colour}}, }{}\href{https://doi.org/10.1007/JHEP08(2015)003}{JHEP \textbf{08}
  (2015) 003}, \href{http://arxiv.org/abs/1505.01681}{{\normalfont\ttfamily
  arXiv:1505.01681}}\relax
\mciteBstWouldAddEndPuncttrue
\mciteSetBstMidEndSepPunct{\mcitedefaultmidpunct}
{\mcitedefaultendpunct}{\mcitedefaultseppunct}\relax
\EndOfBibitem
\bibitem{Greco:2004rm}
V.~Greco and C.~M. Ko,
  \ifthenelse{\boolean{articletitles}}{\emph{{Hadronization via coalescence}},
  }{}\href{https://doi.org/10.1556/APH.24.2005.1-4.32}{Acta Phys.\ Hung.\
  \textbf{A24} (2005) 235},
  \href{http://arxiv.org/abs/nucl-th/0405040}{{\normalfont\ttfamily
  arXiv:nucl-th/0405040}}\relax
\mciteBstWouldAddEndPuncttrue
\mciteSetBstMidEndSepPunct{\mcitedefaultmidpunct}
{\mcitedefaultendpunct}{\mcitedefaultseppunct}\relax
\EndOfBibitem
\bibitem{Armesto:2006ph}
N.~Armesto, \ifthenelse{\boolean{articletitles}}{\emph{{Nuclear shadowing}},
  }{}\href{https://doi.org/10.1088/0954-3899/32/11/R01}{J.\ Phys.\
  \textbf{G32} (2006) R367},
  \href{http://arxiv.org/abs/hep-ph/0604108}{{\normalfont\ttfamily
  arXiv:hep-ph/0604108}}\relax
\mciteBstWouldAddEndPuncttrue
\mciteSetBstMidEndSepPunct{\mcitedefaultmidpunct}
{\mcitedefaultendpunct}{\mcitedefaultseppunct}\relax
\EndOfBibitem
\bibitem{Vitev:2007ve}
I.~Vitev, \ifthenelse{\boolean{articletitles}}{\emph{{Non-Abelian energy loss
  in cold nuclear matter}},
  }{}\href{https://doi.org/10.1103/PhysRevC.75.064906}{Phys.\ Rev.\
  \textbf{C75} (2007) 064906},
  \href{http://arxiv.org/abs/hep-ph/0703002}{{\normalfont\ttfamily
  arXiv:hep-ph/0703002}}\relax
\mciteBstWouldAddEndPuncttrue
\mciteSetBstMidEndSepPunct{\mcitedefaultmidpunct}
{\mcitedefaultendpunct}{\mcitedefaultseppunct}\relax
\EndOfBibitem
\bibitem{PhysRevLett.68.1834}
S.~Gavin and J.~Milana, \ifthenelse{\boolean{articletitles}}{\emph{Energy loss
  at large ${\mathit{x}}_{\mathit{f}}$ in nuclear collisions},
  }{}\href{https://doi.org/10.1103/PhysRevLett.68.1834}{Phys.\ Rev.\ Lett.\
  \textbf{68} (1992) 1834}\relax
\mciteBstWouldAddEndPuncttrue
\mciteSetBstMidEndSepPunct{\mcitedefaultmidpunct}
{\mcitedefaultendpunct}{\mcitedefaultseppunct}\relax
\EndOfBibitem
\bibitem{dai2021parton}
T.~Dai, J.-F. Paquet, D.~Teaney, and S.~A. Bass,
  \ifthenelse{\boolean{articletitles}}{\emph{{Parton energy loss in a hard-soft
  factorized approach}},
  }{}\href{https://doi.org/10.1103/PhysRevC.105.034905}{Phys.\ Rev.\
  \textbf{C105} (2022) 034905},
  \href{http://arxiv.org/abs/2012.03441}{{\normalfont\ttfamily
  arXiv:2012.03441}}\relax
\mciteBstWouldAddEndPuncttrue
\mciteSetBstMidEndSepPunct{\mcitedefaultmidpunct}
{\mcitedefaultendpunct}{\mcitedefaultseppunct}\relax
\EndOfBibitem
\bibitem{breakup2008}
PHENIX collaboration, M.~Wysocki,
  \ifthenelse{\boolean{articletitles}}{\emph{{Measurements of cold nuclear
  matter effects on \jpsi in the PHENIX experiment via deuteron-gold
  collisions}}, }{}\href{https://doi.org/10.1088/0954-3899/35/10/104069}{J.\
  Phys.\  \textbf{G35} (2008) 104069},
  \href{http://arxiv.org/abs/0804.4660}{{\normalfont\ttfamily
  arXiv:0804.4660}}\relax
\mciteBstWouldAddEndPuncttrue
\mciteSetBstMidEndSepPunct{\mcitedefaultmidpunct}
{\mcitedefaultendpunct}{\mcitedefaultseppunct}\relax
\EndOfBibitem
\bibitem{2013aliceheavy}
R.~Grajcarek, \ifthenelse{\boolean{articletitles}}{\emph{Measurement of
  heavy-flavor production in {Pb—Pb} collisions at the {LHC} with {ALICE}},
  }{}\href{https://doi.org/10.1088/1742-6596/420/1/012032}{Journal of Physics:
  Conference Series \textbf{420} (2013) 012032}\relax
\mciteBstWouldAddEndPuncttrue
\mciteSetBstMidEndSepPunct{\mcitedefaultmidpunct}
{\mcitedefaultendpunct}{\mcitedefaultseppunct}\relax
\EndOfBibitem
\bibitem{Lee:2018njd}
Y.-J. Lee, \ifthenelse{\boolean{articletitles}}{\emph{{Quark\textendash{}gluon
  droplets engineered}},
  }{}\href{https://doi.org/10.1038/s41567-018-0375-6}{Nature Physics
  \textbf{15} (2019) 206}\relax
\mciteBstWouldAddEndPuncttrue
\mciteSetBstMidEndSepPunct{\mcitedefaultmidpunct}
{\mcitedefaultendpunct}{\mcitedefaultseppunct}\relax
\EndOfBibitem
\bibitem{RAFELSKI1984215}
J.~Rafelski, \ifthenelse{\boolean{articletitles}}{\emph{Strangeness production
  in the quark gluon plasma},
  }{}\href{https://doi.org/https://doi.org/10.1016/0375-9474(84)90551-7}{Nuclear
  Physics \textbf{A418} (1984) 215}\relax
\mciteBstWouldAddEndPuncttrue
\mciteSetBstMidEndSepPunct{\mcitedefaultmidpunct}
{\mcitedefaultendpunct}{\mcitedefaultseppunct}\relax
\EndOfBibitem
\bibitem{Rafelski:1980fy}
J.~Rafelski, \ifthenelse{\boolean{articletitles}}{\emph{{Extreme states of
  nuclear matter.}},
  }{}\href{https://doi.org/10.1016/0375-9474(82)90260-3}{Nucl.\ Phys.\
  \textbf{374} (1982) 489C}\relax
\mciteBstWouldAddEndPuncttrue
\mciteSetBstMidEndSepPunct{\mcitedefaultmidpunct}
{\mcitedefaultendpunct}{\mcitedefaultseppunct}\relax
\EndOfBibitem
\bibitem{Koch:1984tz}
P.~Koch and J.~Rafelski, \ifthenelse{\boolean{articletitles}}{\emph{{Time
  evolution of strange particle densities in hot hadronic matter}},
  }{}\href{https://doi.org/10.1016/0375-9474(85)90112-5}{Nucl.\ Phys.\
  \textbf{A444} (1985) 678}\relax
\mciteBstWouldAddEndPuncttrue
\mciteSetBstMidEndSepPunct{\mcitedefaultmidpunct}
{\mcitedefaultendpunct}{\mcitedefaultseppunct}\relax
\EndOfBibitem
\bibitem{Koch:1986ud}
P.~Koch, B.~Muller, and J.~Rafelski,
  \ifthenelse{\boolean{articletitles}}{\emph{{Strangeness in relativistic heavy
  ion collisions}},
  }{}\href{https://doi.org/10.1016/0370-1573(86)90096-7}{Phys.\ Rept.\
  \textbf{142} (1986) 167}\relax
\mciteBstWouldAddEndPuncttrue
\mciteSetBstMidEndSepPunct{\mcitedefaultmidpunct}
{\mcitedefaultendpunct}{\mcitedefaultseppunct}\relax
\EndOfBibitem
\bibitem{2017nature}
J.~Adam {\em et~al.}, \ifthenelse{\boolean{articletitles}}{\emph{Enhanced
  production of multi-strange hadrons in high-multiplicity proton–proton
  collisions}, }{}\href{https://doi.org/10.1038/nphys4111}{Nature Physics
  \textbf{13} (2017) 535–539}\relax
\mciteBstWouldAddEndPuncttrue
\mciteSetBstMidEndSepPunct{\mcitedefaultmidpunct}
{\mcitedefaultendpunct}{\mcitedefaultseppunct}\relax
\EndOfBibitem
\bibitem{PHENIX:2013kod}
PHENIX collaboration, A.~Adare {\em et~al.},
  \ifthenelse{\boolean{articletitles}}{\emph{{Spectra and ratios of identified
  particles in Au+Au and $d$+Au collisions at $\sqrt{s_{NN}}=200$ GeV}},
  }{}\href{https://doi.org/10.1103/PhysRevC.88.024906}{Phys.\ Rev.\
  \textbf{C88} (2013) 024906},
  \href{http://arxiv.org/abs/1304.3410}{{\normalfont\ttfamily
  arXiv:1304.3410}}\relax
\mciteBstWouldAddEndPuncttrue
\mciteSetBstMidEndSepPunct{\mcitedefaultmidpunct}
{\mcitedefaultendpunct}{\mcitedefaultseppunct}\relax
\EndOfBibitem
\bibitem{STAR:2006uve}
STAR collaboration, B.~I. Abelev {\em et~al.},
  \ifthenelse{\boolean{articletitles}}{\emph{{Identified baryon and meson
  distributions at large transverse momenta from Au+Au collisions at \sqsnn =
  200 \gev}}, }{}\href{https://doi.org/10.1103/PhysRevLett.97.152301}{Phys.\
  Rev.\ Lett.\  \textbf{97} (2006) 152301},
  \href{http://arxiv.org/abs/nucl-ex/0606003}{{\normalfont\ttfamily
  arXiv:nucl-ex/0606003}}\relax
\mciteBstWouldAddEndPuncttrue
\mciteSetBstMidEndSepPunct{\mcitedefaultmidpunct}
{\mcitedefaultendpunct}{\mcitedefaultseppunct}\relax
\EndOfBibitem
\bibitem{LHCb-PAPER-2014-047}
LHCb collaboration, R.~Aaij {\em et~al.},
  \ifthenelse{\boolean{articletitles}}{\emph{{Precision luminosity measurements
  at LHCb}}, }{}\href{https://doi.org/10.1088/1748-0221/9/12/P12005}{JINST
  \textbf{9} (2014) P12005},
  \href{http://arxiv.org/abs/1410.0149}{{\normalfont\ttfamily
  arXiv:1410.0149}}\relax
\mciteBstWouldAddEndPuncttrue
\mciteSetBstMidEndSepPunct{\mcitedefaultmidpunct}
{\mcitedefaultendpunct}{\mcitedefaultseppunct}\relax
\EndOfBibitem
\bibitem{PDG2022}
Particle Data Group, R.~L. Workman {\em et~al.},
  \ifthenelse{\boolean{articletitles}}{\emph{{\href{http://pdg.lbl.gov/}{Review
  of particle physics}}}, }{}\href{https://doi.org/10.1093/ptep/ptac097}{Prog.\
  Theor.\ Exp.\ Phys.\  \textbf{2022} (2022) 083C01}\relax
\mciteBstWouldAddEndPuncttrue
\mciteSetBstMidEndSepPunct{\mcitedefaultmidpunct}
{\mcitedefaultendpunct}{\mcitedefaultseppunct}\relax
\EndOfBibitem
\bibitem{LHCb-DP-2008-001}
LHCb collaboration, A.~A. Alves~Jr.\ {\em et~al.},
  \ifthenelse{\boolean{articletitles}}{\emph{{The \lhcb detector at the LHC}},
  }{}\href{https://doi.org/10.1088/1748-0221/3/08/S08005}{JINST \textbf{3}
  (2008) S08005}\relax
\mciteBstWouldAddEndPuncttrue
\mciteSetBstMidEndSepPunct{\mcitedefaultmidpunct}
{\mcitedefaultendpunct}{\mcitedefaultseppunct}\relax
\EndOfBibitem
\bibitem{LHCb-DP-2014-002}
LHCb collaboration, R.~Aaij {\em et~al.},
  \ifthenelse{\boolean{articletitles}}{\emph{{LHCb detector performance}},
  }{}\href{https://doi.org/10.1142/S0217751X15300227}{Int.\ J.\ Mod.\ Phys.\
  \textbf{A30} (2015) 1530022},
  \href{http://arxiv.org/abs/1412.6352}{{\normalfont\ttfamily
  arXiv:1412.6352}}\relax
\mciteBstWouldAddEndPuncttrue
\mciteSetBstMidEndSepPunct{\mcitedefaultmidpunct}
{\mcitedefaultendpunct}{\mcitedefaultseppunct}\relax
\EndOfBibitem
\bibitem{LHCb-DP-2014-001}
R.~Aaij {\em et~al.}, \ifthenelse{\boolean{articletitles}}{\emph{{Performance
  of the LHCb Vertex Locator}},
  }{}\href{https://doi.org/10.1088/1748-0221/9/09/P09007}{JINST \textbf{9}
  (2014) P09007}, \href{http://arxiv.org/abs/1405.7808}{{\normalfont\ttfamily
  arXiv:1405.7808}}\relax
\mciteBstWouldAddEndPuncttrue
\mciteSetBstMidEndSepPunct{\mcitedefaultmidpunct}
{\mcitedefaultendpunct}{\mcitedefaultseppunct}\relax
\EndOfBibitem
\bibitem{LHCb-DP-2017-001}
P.~d'Argent {\em et~al.}, \ifthenelse{\boolean{articletitles}}{\emph{{Improved
  performance of the LHCb Outer Tracker in LHC Run 2}},
  }{}\href{https://doi.org/10.1088/1748-0221/12/11/P11016}{JINST \textbf{12}
  (2017) P11016}, \href{http://arxiv.org/abs/1708.00819}{{\normalfont\ttfamily
  arXiv:1708.00819}}\relax
\mciteBstWouldAddEndPuncttrue
\mciteSetBstMidEndSepPunct{\mcitedefaultmidpunct}
{\mcitedefaultendpunct}{\mcitedefaultseppunct}\relax
\EndOfBibitem
\bibitem{LHCb-DP-2012-003}
M.~Adinolfi {\em et~al.},
  \ifthenelse{\boolean{articletitles}}{\emph{{Performance of the \lhcb RICH
  detector at the LHC}},
  }{}\href{https://doi.org/10.1140/epjc/s10052-013-2431-9}{Eur.\ Phys.\ J.\
  \textbf{C73} (2013) 2431},
  \href{http://arxiv.org/abs/1211.6759}{{\normalfont\ttfamily
  arXiv:1211.6759}}\relax
\mciteBstWouldAddEndPuncttrue
\mciteSetBstMidEndSepPunct{\mcitedefaultmidpunct}
{\mcitedefaultendpunct}{\mcitedefaultseppunct}\relax
\EndOfBibitem
\bibitem{LHCb-DP-2012-002}
A.~A. Alves~Jr.\ {\em et~al.},
  \ifthenelse{\boolean{articletitles}}{\emph{{Performance of the LHCb muon
  system}}, }{}\href{https://doi.org/10.1088/1748-0221/8/02/P02022}{JINST
  \textbf{8} (2013) P02022},
  \href{http://arxiv.org/abs/1211.1346}{{\normalfont\ttfamily
  arXiv:1211.1346}}\relax
\mciteBstWouldAddEndPuncttrue
\mciteSetBstMidEndSepPunct{\mcitedefaultmidpunct}
{\mcitedefaultendpunct}{\mcitedefaultseppunct}\relax
\EndOfBibitem
\bibitem{LHCb-DP-2012-004}
R.~Aaij {\em et~al.}, \ifthenelse{\boolean{articletitles}}{\emph{{The \lhcb
  trigger and its performance in 2011}},
  }{}\href{https://doi.org/10.1088/1748-0221/8/04/P04022}{JINST \textbf{8}
  (2013) P04022}, \href{http://arxiv.org/abs/1211.3055}{{\normalfont\ttfamily
  arXiv:1211.3055}}\relax
\mciteBstWouldAddEndPuncttrue
\mciteSetBstMidEndSepPunct{\mcitedefaultmidpunct}
{\mcitedefaultendpunct}{\mcitedefaultseppunct}\relax
\EndOfBibitem
\bibitem{Sjostrand:2007gs}
T.~Sj\"{o}strand, S.~Mrenna, and P.~Skands,
  \ifthenelse{\boolean{articletitles}}{\emph{{A brief introduction to PYTHIA
  8.1}}, }{}\href{https://doi.org/10.1016/j.cpc.2008.01.036}{Comput.\ Phys.\
  Commun.\  \textbf{178} (2008) 852},
  \href{http://arxiv.org/abs/0710.3820}{{\normalfont\ttfamily
  arXiv:0710.3820}}\relax
\mciteBstWouldAddEndPuncttrue
\mciteSetBstMidEndSepPunct{\mcitedefaultmidpunct}
{\mcitedefaultendpunct}{\mcitedefaultseppunct}\relax
\EndOfBibitem
\bibitem{EPOS}
T.~Pierog {\em et~al.}, \ifthenelse{\boolean{articletitles}}{\emph{{EPOS LHC:
  Test of collective hadronization with data measured at the CERN Large Hadron
  Collider}}, }{}\href{https://doi.org/10.1103/PhysRevC.92.034906}{Phys.\ Rev.\
   \textbf{C92} (2015) 034906},
  \href{http://arxiv.org/abs/1306.0121}{{\normalfont\ttfamily
  arXiv:1306.0121}}\relax
\mciteBstWouldAddEndPuncttrue
\mciteSetBstMidEndSepPunct{\mcitedefaultmidpunct}
{\mcitedefaultendpunct}{\mcitedefaultseppunct}\relax
\EndOfBibitem
\bibitem{Lange:2001uf}
D.~J. Lange, \ifthenelse{\boolean{articletitles}}{\emph{{The EvtGen particle
  decay simulation package}},
  }{}\href{https://doi.org/10.1016/S0168-9002(01)00089-4}{Nucl.\ Instrum.\
  Meth.\  \textbf{A462} (2001) 152}\relax
\mciteBstWouldAddEndPuncttrue
\mciteSetBstMidEndSepPunct{\mcitedefaultmidpunct}
{\mcitedefaultendpunct}{\mcitedefaultseppunct}\relax
\EndOfBibitem
\bibitem{Agostinelli:2002hh}
Geant4 collaboration, S.~Agostinelli {\em et~al.},
  \ifthenelse{\boolean{articletitles}}{\emph{{Geant4: A simulation toolkit}},
  }{}\href{https://doi.org/10.1016/S0168-9002(03)01368-8}{Nucl.\ Instrum.\
  Meth.\  \textbf{A506} (2003) 250}\relax
\mciteBstWouldAddEndPuncttrue
\mciteSetBstMidEndSepPunct{\mcitedefaultmidpunct}
{\mcitedefaultendpunct}{\mcitedefaultseppunct}\relax
\EndOfBibitem
\bibitem{Allison:2006ve}
Geant4 collaboration, J.~Allison {\em et~al.},
  \ifthenelse{\boolean{articletitles}}{\emph{{Geant4 developments and
  applications}}, }{}\href{https://doi.org/10.1109/TNS.2006.869826}{IEEE
  Trans.\ Nucl.\ Sci.\  \textbf{53} (2006) 270}\relax
\mciteBstWouldAddEndPuncttrue
\mciteSetBstMidEndSepPunct{\mcitedefaultmidpunct}
{\mcitedefaultendpunct}{\mcitedefaultseppunct}\relax
\EndOfBibitem
\bibitem{LHCb-PROC-2015-011}
G.~Dujany and B.~Storaci, \ifthenelse{\boolean{articletitles}}{\emph{{Real-time
  alignment and calibration of the LHCb Detector in Run II}},
  }{}\href{https://doi.org/10.1088/1742-6596/664/8/082010}{J.\ Phys.\ Conf.\
  Ser.\  \textbf{664} (2015) 082010}\relax
\mciteBstWouldAddEndPuncttrue
\mciteSetBstMidEndSepPunct{\mcitedefaultmidpunct}
{\mcitedefaultendpunct}{\mcitedefaultseppunct}\relax
\EndOfBibitem
\bibitem{LHCb-DP-2016-001}
R.~Aaij {\em et~al.}, \ifthenelse{\boolean{articletitles}}{\emph{{Tesla: an
  application for real-time data analysis in High Energy Physics}},
  }{}\href{https://doi.org/10.1016/j.cpc.2016.07.022}{Comput.\ Phys.\ Commun.\
  \textbf{208} (2016) 35},
  \href{http://arxiv.org/abs/1604.05596}{{\normalfont\ttfamily
  arXiv:1604.05596}}\relax
\mciteBstWouldAddEndPuncttrue
\mciteSetBstMidEndSepPunct{\mcitedefaultmidpunct}
{\mcitedefaultendpunct}{\mcitedefaultseppunct}\relax
\EndOfBibitem
\bibitem{Skwarnicki:1986xj}
T.~Skwarnicki, {\em {A study of the radiative cascade transitions between the
  Upsilon-prime and Upsilon resonances}}, PhD thesis, Institute of Nuclear
  Physics, Krakow, 1986,
  {\href{http://inspirehep.net/record/230779/}{DESY-F31-86-02}}\relax
\mciteBstWouldAddEndPuncttrue
\mciteSetBstMidEndSepPunct{\mcitedefaultmidpunct}
{\mcitedefaultendpunct}{\mcitedefaultseppunct}\relax
\EndOfBibitem
\bibitem{Pivk:2004ty}
M.~Pivk and F.~R. Le~Diberder,
  \ifthenelse{\boolean{articletitles}}{\emph{{sPlot: A statistical tool to
  unfold data distributions}},
  }{}\href{https://doi.org/10.1016/j.nima.2005.08.106}{Nucl.\ Instrum.\ Meth.\
  \textbf{A555} (2005) 356},
  \href{http://arxiv.org/abs/physics/0402083}{{\normalfont\ttfamily
  arXiv:physics/0402083}}\relax
\mciteBstWouldAddEndPuncttrue
\mciteSetBstMidEndSepPunct{\mcitedefaultmidpunct}
{\mcitedefaultendpunct}{\mcitedefaultseppunct}\relax
\EndOfBibitem
\bibitem{BukinF}
A.~D. Bukin, \ifthenelse{\boolean{articletitles}}{\emph{{Fitting function for
  asymmetric peaks}},
  }{}\href{http://arxiv.org/abs/0711.4449}{{\normalfont\ttfamily
  arXiv:0711.4449}}\relax
\mciteBstWouldAddEndPuncttrue
\mciteSetBstMidEndSepPunct{\mcitedefaultmidpunct}
{\mcitedefaultendpunct}{\mcitedefaultseppunct}\relax
\EndOfBibitem
\bibitem{LHCb-PUB-2016-021}
L.~Anderlini {\em et~al.}, \ifthenelse{\boolean{articletitles}}{\emph{{The
  PIDCalib package}}, }{}
  \href{http://cdsweb.cern.ch/search?p=LHCb-PUB-2016-021&f=reportnumber&action_search=Search&c=LHCb+Notes}
  {LHCb-PUB-2016-021}, 2016\relax
\mciteBstWouldAddEndPuncttrue
\mciteSetBstMidEndSepPunct{\mcitedefaultmidpunct}
{\mcitedefaultendpunct}{\mcitedefaultseppunct}\relax
\EndOfBibitem
\bibitem{Lansberg:2016deg}
J.-P. Lansberg and H.-S. Shao,
  \ifthenelse{\boolean{articletitles}}{\emph{{Towards an automated tool to
  evaluate the impact of the nuclear modification of the gluon density on
  quarkonium, D and B meson production in proton–nucleus collisions}},
  }{}\href{https://doi.org/10.1140/epjc/s10052-016-4575-x}{Eur.\ Phys.\ J.\
  \textbf{C77} (2017) 1},
  \href{http://arxiv.org/abs/1610.05382}{{\normalfont\ttfamily
  arXiv:1610.05382}}\relax
\mciteBstWouldAddEndPuncttrue
\mciteSetBstMidEndSepPunct{\mcitedefaultmidpunct}
{\mcitedefaultendpunct}{\mcitedefaultseppunct}\relax
\EndOfBibitem
\bibitem{Kusina:2017gkz}
A.~Kusina, J.-P. Lansberg, I.~Schienbein, and H.-S. Shao,
  \ifthenelse{\boolean{articletitles}}{\emph{{Gluon shadowing in heavy-flavor
  production at the LHC}},
  }{}\href{https://doi.org/10.1103/PhysRevLett.121.052004}{Phys.\ Rev.\ Lett.\
  \textbf{121} (2018) 052004},
  \href{http://arxiv.org/abs/1712.07024}{{\normalfont\ttfamily
  arXiv:1712.07024}}\relax
\mciteBstWouldAddEndPuncttrue
\mciteSetBstMidEndSepPunct{\mcitedefaultmidpunct}
{\mcitedefaultendpunct}{\mcitedefaultseppunct}\relax
\EndOfBibitem
\bibitem{Kusina:2020dki}
A.~Kusina, J.-P. Lansberg, I.~Schienbein, and H.-S. Shao,
  \ifthenelse{\boolean{articletitles}}{\emph{{Reweighted nuclear PDFs using
  heavy-flavor production data at the LHC}},
  }{}\href{https://doi.org/10.1103/PhysRevD.104.014010}{Phys.\ Rev.\
  \textbf{D104} (2021) 014010},
  \href{http://arxiv.org/abs/2012.11462}{{\normalfont\ttfamily
  arXiv:2012.11462}}\relax
\mciteBstWouldAddEndPuncttrue
\mciteSetBstMidEndSepPunct{\mcitedefaultmidpunct}
{\mcitedefaultendpunct}{\mcitedefaultseppunct}\relax
\EndOfBibitem
\bibitem{Shao:2012iz}
H.-S. Shao, \ifthenelse{\boolean{articletitles}}{\emph{{HELAC-Onia: An
  automatic matrix element generator for heavy quarkonium physics}},
  }{}\href{https://doi.org/10.1016/j.cpc.2013.05.023}{Comput.\ Phys.\ Commun.\
  \textbf{184} (2013) 2562},
  \href{http://arxiv.org/abs/1212.5293}{{\normalfont\ttfamily
  arXiv:1212.5293}}\relax
\mciteBstWouldAddEndPuncttrue
\mciteSetBstMidEndSepPunct{\mcitedefaultmidpunct}
{\mcitedefaultendpunct}{\mcitedefaultseppunct}\relax
\EndOfBibitem
\bibitem{Shao:2015vga}
H.-S. Shao, \ifthenelse{\boolean{articletitles}}{\emph{{HELAC-Onia 2.0: An
  upgraded matrix-element and event generator for heavy quarkonium physics}},
  }{}\href{https://doi.org/10.1016/j.cpc.2015.09.011}{Comput.\ Phys.\ Commun.\
  \textbf{198} (2016) 238},
  \href{http://arxiv.org/abs/1507.03435}{{\normalfont\ttfamily
  arXiv:1507.03435}}\relax
\mciteBstWouldAddEndPuncttrue
\mciteSetBstMidEndSepPunct{\mcitedefaultmidpunct}
{\mcitedefaultendpunct}{\mcitedefaultseppunct}\relax
\EndOfBibitem
\bibitem{Eskola:2016oht}
K.~J. Eskola, P.~Paakkinen, H.~Paukkunen, and C.~A. Salgado,
  \ifthenelse{\boolean{articletitles}}{\emph{{EPPS16: Nuclear parton
  distributions with LHC data}},
  }{}\href{https://doi.org/10.1140/epjc/s10052-017-4725-9}{Eur.\ Phys.\ J.\
  \textbf{C77} (2017) 163},
  \href{http://arxiv.org/abs/1612.05741}{{\normalfont\ttfamily
  arXiv:1612.05741}}\relax
\mciteBstWouldAddEndPuncttrue
\mciteSetBstMidEndSepPunct{\mcitedefaultmidpunct}
{\mcitedefaultendpunct}{\mcitedefaultseppunct}\relax
\EndOfBibitem
\bibitem{LHCb-PAPER-2022-007}
LHCb collaboration, R.~Aaij {\em et~al.},
  \ifthenelse{\boolean{articletitles}}{\emph{{Measurement of the prompt $\Dz$
  nuclear modification factor in $\proton$Pb collisions at $\sqrt{s_{NN}}$ =
  8.16 TeV}}, }{}\href{https://doi.org/10.1103/PhysRevLett.131.102301}{Phys.\
  Rev.\ Lett.\  \textbf{131} (2023) 102301},
  \href{http://arxiv.org/abs/2205.03936}{{\normalfont\ttfamily
  arXiv:2205.03936}}\relax
\mciteBstWouldAddEndPuncttrue
\mciteSetBstMidEndSepPunct{\mcitedefaultmidpunct}
{\mcitedefaultendpunct}{\mcitedefaultseppunct}\relax
\EndOfBibitem
\end{mcitethebibliography}

\newpage
\centerline
{\large\bf LHCb collaboration}
\begin
{flushleft}
\small
R.~Aaij$^{32}$\lhcborcid{0000-0003-0533-1952},
A.S.W.~Abdelmotteleb$^{50}$\lhcborcid{0000-0001-7905-0542},
C.~Abellan~Beteta$^{44}$,
F.~Abudin{\'e}n$^{50}$\lhcborcid{0000-0002-6737-3528},
T.~Ackernley$^{54}$\lhcborcid{0000-0002-5951-3498},
B.~Adeva$^{40}$\lhcborcid{0000-0001-9756-3712},
M.~Adinolfi$^{48}$\lhcborcid{0000-0002-1326-1264},
P.~Adlarson$^{77}$\lhcborcid{0000-0001-6280-3851},
H.~Afsharnia$^{9}$,
C.~Agapopoulou$^{13}$\lhcborcid{0000-0002-2368-0147},
C.A.~Aidala$^{78}$\lhcborcid{0000-0001-9540-4988}
% ...

{\footnotesize \it

% ...
$^{9}$Universit{\'e} Clermont Auvergne, CNRS/IN2P3, LPC, Clermont-Ferrand, France\\
% ...
$^{13}$LPNHE, Sorbonne Universit{\'e}, Paris Diderot Sorbonne Paris Cit{\'e}, CNRS/IN2P3, Paris, France\\
% ...
$^{32}$Nikhef National Institute for Subatomic Physics, Amsterdam, Netherlands\\
% ...
$^{40}$Instituto Galego de F{\'\i}sica de Altas Enerx{\'\i}as (IGFAE), Universidade de Santiago de Compostela, Santiago de Compostela, Spain\\
% ...
$^{44}$Physik-Institut, Universit{\"a}t Z{\"u}rich, Z{\"u}rich, Switzerland\\
% ...
$^{48}$H.H. Wills Physics Laboratory, University of Bristol, Bristol, United Kingdom\\
% ...
$^{50}$Department of Physics, University of Warwick, Coventry, United Kingdom\\
% ...
$^{53}$School of Physics and Astronomy, University of Glasgow, Glasgow, United Kingdom\\
$^{54}$Oliver Lodge Laboratory, University of Liverpool, Liverpool, United Kingdom\\
% ...
$^{62}$Syracuse University, Syracuse, NY, United States\\
% ...
$^{77}$Department of Physics and Astronomy, Uppsala University, Uppsala, Sweden, associated to $^{53}$\\
$^{78}$University of Michigan, Ann Arbor, MI, United States, associated to $^{62}$\\
\bigskip
% ...
}
\end{flushleft}

%
%
%
%

\end{document}